\documentclass{aa}  
\graphicspath{}
\usepackage{chngcntr}
\usepackage[overload]{empheq}
\usepackage{tabularx}
\usepackage[11pt]{moresize}
\usepackage{xtab,afterpage}
\usepackage{booktabs}                     
\usepackage[export]{adjustbox}
\usepackage[singlelinecheck=false]{caption}
\usepackage{subcaption}
\usepackage{subdepth}
\usepackage{amsmath}
\let\Sigma\varSigma
\let\Psi\varPsi
\usepackage{fixltx2e} 
\usepackage{nameref}
\usepackage{varioref}
\usepackage{xcolor}
\usepackage{hyperref}
\hypersetup{
    colorlinks,
    linkcolor={red!50!black},
    citecolor={blue!80!black},
    urlcolor={blue!80!black}
}
\usepackage{textcomp}
\usepackage{cleveref}
\usepackage[utf8]{inputenc}
\usepackage{natbib}
\bibpunct{(}{)}{;}{a}{}{,} 
\counterwithout{figure}{section}
\usepackage{cases}
\usepackage{graphicx}
\usepackage{placeins}
\usepackage{upgreek}
\usepackage{amsmath}	
\usepackage{txfonts}
\begin{document} 
\setlength{\abovedisplayskip}{12pt}
\setlength{\belowdisplayskip}{12pt}

\title{The ALMA-PILS survey: 3D modeling of the envelope, disks and dust filament of IRAS~16293--2422}

   \author{S.~K.~Jacobsen
          \inst{\ref{nbi-starplan}}
          \and
          J.~K.~J\o rgensen\inst{\ref{nbi-starplan}} 
	    \and 
          M.~H.~D.~van der Wiel\inst{\ref{nbi-starplan},\ref{astron}}
        \and
        H.~Calcutt\inst{\ref{nbi-starplan}}
        \and
        T.~L.~Bourke\inst{\ref{SKA}}
        \and
        C.~Brinch\inst{\ref{NBIA}}
	    \and
          A.~Coutens\inst{\ref{bord},\ref{ucl}}
	    \and
          M.~N.~Drozdovskaya\inst{\ref{bern}}
          \and
          L.~E.~Kristensen\inst{\ref{nbi-starplan}}
          \and
          H.~S.~P.~M{\"u}ller\inst{\ref{koln}}
          \and
	    S.~F.~Wampfler\inst{\ref{bern}}
          }

 \institute{
     Centre for Star and Planet Formation, Niels Bohr Institute and Natural History Museum of Denmark, University of Copenhagen, {\O}ster Voldgade 5--7, DK-1350 Copenhagen K, Denmark
              \\
		  \email{steffen.jacobsen@nbi.ku.dk}
               \label{nbi-starplan}
    \and
    ASTRON, the Netherlands Institute for Radio Astronomy, Postbus 2, 7990 AA Dwingeloo, The Netherlands
    \label{astron}
    \and
    SKA Organization, Jodrell Bank Observatory, Lower Withington, Macclesfield, Cheshire SK11 9DL, UK
    \label{SKA}
    \and
    Niels Bohr International Academy, Niels Bohr Institute, University of Copenhagen, Blegdamsvej 17, 2100 Copenhagen {\O}, Denmark
    \label{NBIA}
    \and
    Laboratoire d'astrophysique de Bordeaux, Univ. Bordeaux, CNRS, B18N, all{\'e}e Geoffroy Saint-Hilaire, 33615 Pessac, France
    \label{bord}
    \and
    Department of Physics and Astronomy, University College London, Gower St., London, WC1E 6BT, UK
    \label{ucl}
    \and
    Center for Space and Habitability (CSH), Universität Bern, Sidlerstrasse 5, CH-3012 Bern, Switzerland
    \label{bern}
    \and
    I.~Physikalisches Institut, Universit{\"a}t zu K{\"o}ln, Z{\"u}lpicher Str. 77, 50937 K{\"o}ln, Germany
    \label{koln}
}
    \date{Received 28/07/2017; accepted 02/12/2017}

 
  \abstract
  {\textit{Context.} The Class~0 protostellar binary IRAS~16293--2422 is an interesting target for (sub)millimeter observations due to, both, the rich chemistry toward the two main components of the binary and its complex morphology. Its proximity to Earth allows the study of its physical and chemical structure on solar system scales using high angular resolution observations. Such data reveal a complex morphology that cannot be accounted for in traditional, spherical 1D models of the envelope.

\textit{Aims.} The purpose of this paper is to study the environment of the two components of the binary through 3D radiative transfer modeling and to compare with data from the Atacama Large Millimeter/submillimeter Array. Such comparisons can be used to constrain the protoplanetary disk structures, the luminosities of the two components of the binary and the chemistry of simple species.

\textit{Methods.} We present $^{13}$CO, C$^{17}$O and C$^{18}$O $J$=3--2 observations from the ALMA Protostellar Interferometric Line Survey (PILS), together with a qualitative study of the dust and gas density distribution of IRAS 16293-2422. A 3D dust and gas model including disks and a dust filament between the two protostars is constructed which qualitatively reproduces the dust continuum and gas line emission.

\textit{Results.} Radiative transfer modeling in our sampled parameter space suggests that, while the disk around source~A could not be constrained, the disk around source~B has to be vertically extended. This puffed-up structure can be obtained with both a protoplanetary disk model with an unexpectedly high scale-height and with the density solution from an infalling, rotating collapse. Combined constraints on our 3D model, from observed dust continuum and CO isotopologue emission between the sources, corroborate that source~A should be at least six times more luminous than source~B. We also demonstrate that the volume of high-temperature regions where complex organic molecules arise is sensitive to whether or not the total luminosity is in a single radiation source or distributed into two sources, affecting the interpretation of earlier chemical modeling efforts of the IRAS 16293-2422 hot corino which used a single-source approximation.

\textit{Conclusions.} Radiative transfer modeling of source~A and B, with the density solution of an infalling, rotating collapse or a protoplanetary disk model, can match the constraints for the disk-like emission around source~A and B from the observed dust continuum and CO isotopologue gas emission. If a protoplanetary disk model is used around source~B, it has to have an unusually high scale-height in order to reach the dust continuum peak emission value, while fulfilling the other observational constraints. Our 3D model requires source~A to be much more luminous than source~B; $L_\mathrm{A}$~$\sim$~18~L$_\sun$ and $L_\mathrm{B}$~$\sim$~3~L$_\sun$.
}

   \maketitle
\section{Introduction}
Stars form inside the dense cores of interstellar cold clouds of gas and dust. During the gravitational collapse of the dense core material, the slow rotation of the core itself will eventually cause a protoplanetary disk to appear, due to the conservation of angular momentum \mbox{\citep{1981Icar...48..353C}}. 
When exactly such disks are formed and how they evolve through their early stages remain poorly understood \citep[see, e.g.,][for a review]{2014prpl.conf..173L}.
Well established disks around T~Tauri stars or Class~II young stellar objects have been widely reported and studied extensively \mbox{\citep[e.g., see][]{2009ApJ...700.1502A, 2010A&A...512A..15R}}.
Due to the low spatial resolution of the observations used in earlier studies, the emission from the disks in Class~0 and Class~I sources have been mixed with the emission from the massive amounts of dust in the ambient protostellar envelope, making detection and characterization of disks in Class~0 sources very difficult \citep{2005ApJ...632..973J, 2008ApJ...680..474C, 2009A&A...507..861J}, with only a small number of Class 0 disks with indicated Keplerian rotation being detected so far
\citep{2012Natur.492...83T, 2013A&A...560A.103M,2014ApJ...796..131O, 2014A&A...566A..74L,2014A&A...568L...5C}.
\mbox{\citet{2009A&A...507..861J}} detect compact mm continuum emission around several Class~0 objects, but it is unknown whether these dust components represent rotationally supported Keplerian disks. The exact transition from early density structures to a rotationally supported Keplerian disk is still uncertain, though some tentative observations of the so-called centrifugal barrier, where the infalling, rotating envelope transitions into a 
rotationally supported disk, have been made \mbox{\citep{2014ApJ...791L..38S, 2016ApJ...824...88O}}.
These inner regions around embedded protostars are also particularly interesting from a chemical point of view. Observational studies of Class 0 objects have revealed complex organic molecules \citep[e.g.,][]{2004ApJ...617L..69B, 2004ApJ...616L..27K,2004ApJ...615..354B,2005ApJ...632..973J, 2008A&A...488..959B,2011A&A...534A.100J,2014A&A...563L...2M}, as well as active chemistry in the so-called hot corino region, where water and organic molecules are sublimated into the gas-phase from the dust grain surface \mbox{\citep{2002A&A...390.1001S}}. 
\\
A particularly interesting source in this context is IRAS~16293--2422 (hereafter IRAS~16293), a nearby binary protostar in the $\rho$ Ophiuchus cloud complex, composed of at least two components, IRAS~16293A and IRAS~16293B \mbox{\citep{1989ApJ...337..858W, 1992ApJ...385..306M, 2000ApJ...529..477L}}. IRAS~16293 is one of the most well-studied low-mass protostars, particularly in terms of astrochemistry; IRAS~16293 was the first low-mass star for which complex organics  \mbox{\citep{1995ApJ...447..760V}} and prebiotic molecules \mbox{\citep{2012ApJ...757L...4J}} were detected. 
The source IRAS~16293A appears to be highly active with several detected outflows, while IRAS~16293B is more quiescent, with no known outflows \citep[e.g.,][]{2005ApJ...632..371C, 2013A&A...549L...6K}. 
\citet{2012A&A...542A..14A} find that the magnetic field strengths in IRAS~16293 are comparable to the outflow ram pressure, and thus that the magnetic field is dynamically important for this system.
Both components are embedded within a larger, circumbinary envelope of dust and gas, which has been well studied in previous works using single-dish data and 1D radiative transfer modeling \citep{2002A&A...390.1001S, 2005ApJ...631L..77J, 2010A&A...519A..65C}. 
These models all use a single central radiation source, where disks or complex structures, such as a dust and gas filament, have never been included in the radiative transfer calculations. 
The single-star approximation is not a problem at large distances from the interbinary region as the two sources can then be approximated as a single source, especially if one source is more luminous than the other. 
In contrast, in the interbinary region, two sources are needed in order to estimate the temperature structure. The total luminosity is $L$~=~21~$\pm$~5 L$_{\sun}$ \mbox{\citep{2016A&A...595A.117J}}, while the luminosity ratio of the two protostars is unknown.
IRAS~16293 has been, and continues to be, under intense observational scrutiny due to its rich chemistry and any chemical modeling must assume the parameters of the physical environment in terms of dust and H$_2$ density, temperature structure and the local radiation field. An assessment of the individual luminosities of the sources in IRAS~16293 is crucial for chemical modeling, especially for warm gas-phase chemistry in the hot corinos.

In this paper, we construct a 3D model, including disks, a filament and two radiation sources using high angular resolution data from the ALMA Protostellar Interferometric Line Survey (PILS) as observational constraints. The impact of the binary nature of IRAS~16293 on the temperature distribution is evaluated as opposed to that of a single star. As stellar binaries are abundant, especially in the Class 0 stage \mbox{\citep{2016ApJ...818...73T}}, this temperature evaluation of IRAS~16293 is useful as a case study on the temperature distribution in the environment of protostellar binaries in wide orbits, found in large fractions of star-forming regions \mbox{\citep[e.g.,][]{2004A&A...427..651D}}, as well as a retrospective analysis on the validity of previous works approximating a binary as a single protostar in their radiative transfer modeling, which will be useful for data-mining. 
This work is part of a new generation of complex 2D and 3D radiative transfer models describing Class~0 YSOs \citep[see e.g.,][]{2017ApJ...835..259Y}, which were previously modeled with 1D envelope models.

After the observations and the general radiative transfer setup are described in Sections \ref{sect:obs} and \ref{sect:rad_transf}, the paper is divided into three parts: Section \ref{sec:temp_struct} supplies a general investigation of temperature structures in Class~0 objects and the difference between having one or two central radiation sources in a 1D model of a collapsing envelope. 
Section \ref{sec:dustmodIRAS} shows a 3D model of the dust density structure of IRAS~16293, where we used a dust continuum image at 868 $\upmu$m and the spectral energy distribution (SED) as constraints. Finally, new ALMA-PILS observations of the CO isotopologues $^{13}$CO, C$^{17}$O and C$^{18}$O are used in Section \ref{sec:linemod} to further constrain the physical conditions in the innermost region of IRAS~16293, ending with an evaluation of the nature of the disks, their vertical distribution and the luminosity ratio between the sources, offering a significant increase in physical complexity from earlier models as well as the first luminosity evaluation of the individual sources. 
\section{Observations}
\label{sect:obs}
The PILS program (PI: Jes K. J\o rgensen, project-id: 2013.1.00278.S) is an unbiased spectral survey of IRAS~16293, covering a significant part of ALMA's Band 7, corresponding to a wavelength of roughly 0.8 mm, as well as selected windows of ALMA's Bands 3 and 6, corresponding to $\sim$3~mm and $\sim$1.3~mm, respectively. The Band 7 data cover the frequency range from 329.15 to 362.90~GHz with $\sim$0.2~km~s$^{-1}$ spectral resolution and $\sim$0.5$''$ angular resolution (60~AU beam diameter, assuming $d$=120~pc). The pointing center was set between source~A and B at 
$\alpha_{\mathrm{J2000}}$ = 16$^\mathrm{h}$32$^\mathrm{m}$22$\fs$72;
$\delta_\mathrm{J2000}$ = --24$^{\circ}$28$^{\prime}$34$\farcs$3. Further details about the survey, data reduction and an overview of the observations can be found in \mbox{\citet{2016A&A...595A.117J}}.

\begin{figure}[ht]
  \centering
      \includegraphics[height = 0.3\textheight, width=0.5\textwidth, keepaspectratio]{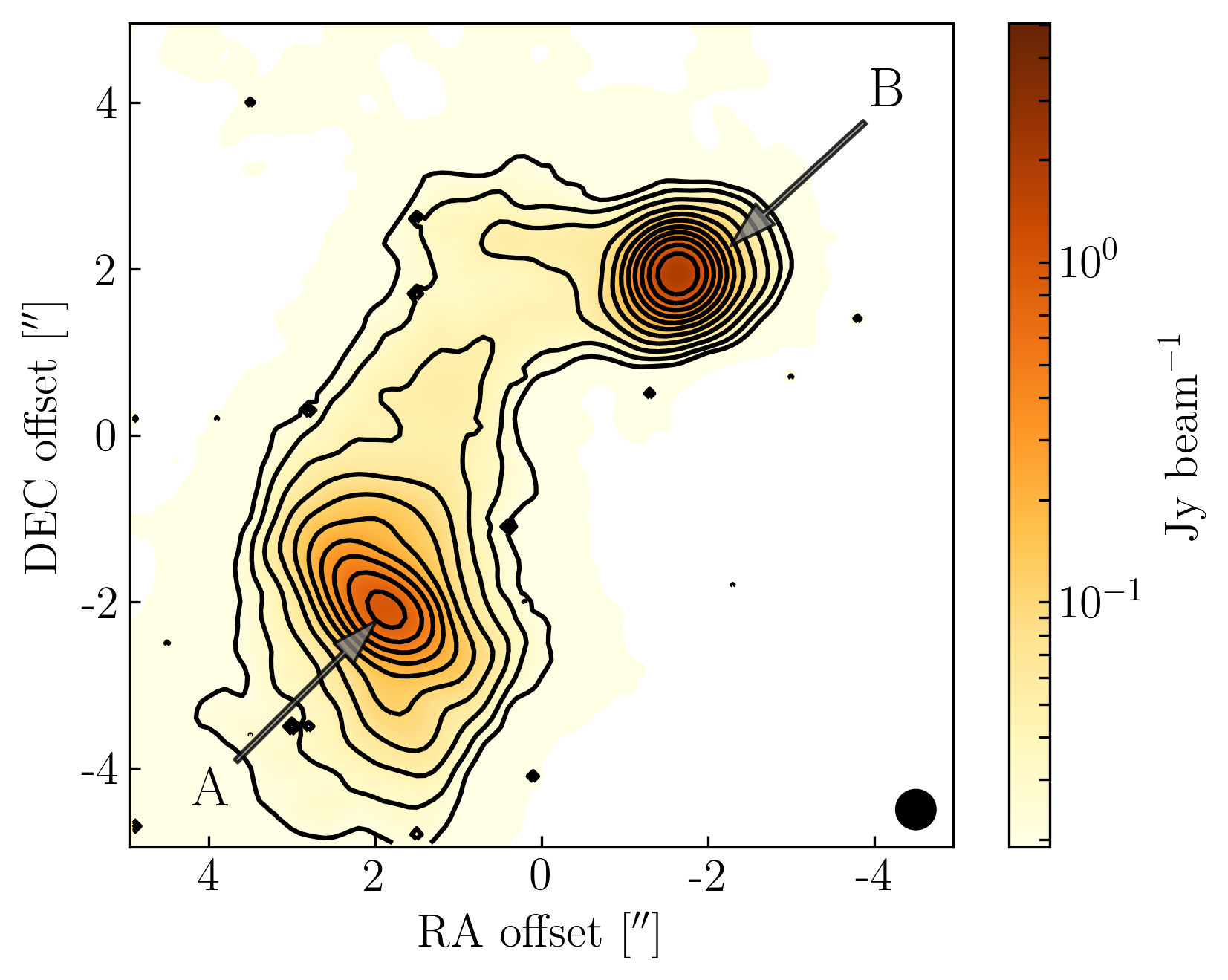}   
      \caption{Dust continuum at 868 $\upmu$m, in log-scale. Contour levels are logarithmically divided between 0.02 and 2 Jy beam$^{-1}$. The RA and DEC offsets are relative to the phase center of the observations. Beamsize is shown in bottom right corner.}
	\label{fig:dust_cont}
\end{figure}

In this paper, we focused on the 868 $\upmu$m continuum and CO isotopologue data, which include data from both the array of 12~m telescopes and the Atacama Compact Array (ACA), allowing both compact and extended structures to be detected. The continuum image was taken as an average of the Band 7 continuum in each channel \mbox{\citep{2016A&A...595A.117J}}. The 868 $\upmu$m dust continuum shows broad extended emission, enveloping both sources. Around each source, emission structures consistent with disks are observed. Source A appears to host a disk close to edge-on, while the emission around source~B appears more circular, that is, suggestive of a face-on disk. 
The continuum is much brighter toward source~B, peaking at 2.0 Jy~beam$^{-1}$, while source~A has a more extended and weaker continuum, peaking at 1.0 Jy~beam$^{-1}$ (Fig. \ref{fig:dust_cont}). 
The dust continuum images from Bands 3 and 6 were not used in this work due to lower spatial resolutions than the Band 7 data, which confuses the disk structures with the filament. 

The gas line emission from the $J$=3--2 transitions of three CO isotopologues, $^{13}$CO \citep[330.588 GHz,][]{klapper_sub-doppler_2000}, C$^{18}$O \citep[329.331 GHz,][]{klapper_sub-doppler_2001} and C$^{17}$O \citep[337.061 GHz,][]{2003ApJ...582..262K} were extracted from the dataset, using the entire pixel map, after the lines were identified. The source~A LSR velocity of 3.1 km~s$^{-1}$ \mbox{\citep{2011A&A...534A.100J}} was used for all lines.
Channels within $\pm$~5~km~s$^{-1}$ of each line were combined into an integrated intensity (moment zero) map for each isotopologue, using the image analysis software \texttt{MIRIAD}. Using this approach, the vast majority of the observed line emission was extracted, while avoiding major line-blending from other species.

\subsection*{Integrated emission maps}
\begin{figure*}[ht]
  \centering
      \includegraphics[height = 0.25\textheight, width=1.0\textwidth, keepaspectratio]{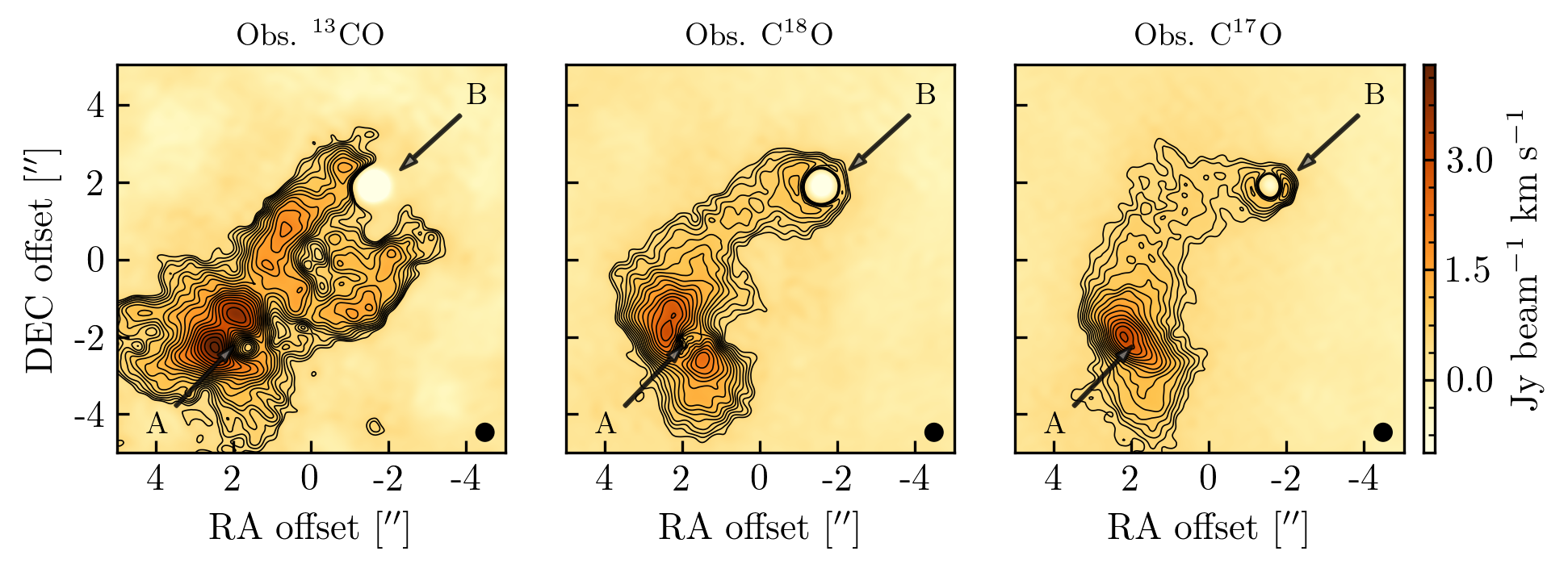}   
      \caption{Zeroth moment maps of the $^{13}$CO, C$^{18}$O, and C$^{17}$O isotopologues from the PILS observations. Contour levels are divided logarithmically from 0.5 to 7.3 Jy beam$^{-1}$km~s$^{-1}$. The RA and DEC offsets are relative to the phase center of the observations.}
      \label{fig:co_obs}   
\end{figure*}

Figure \ref{fig:co_obs} shows the integrated intensity maps of the three isotopologues. C$^{17}$O appears to trace the dust distribution fairly well, as does the C$^{18}$O emission. $^{13}$CO, however, shows a markedly different emission structure, which is likely attributable to optical depth effects, where only the outer layer of the $^{13}$CO structure is probed. Absorption toward source~B is evident in all lines but to a smaller extent in C$^{17}$O, revealing that C$^{17}$O is the line with the lowest optical depth, as expected since it has the lowest abundance \citep{1994ARA&A..32..191W}. An interesting feature is the broad emission arc between the sources, which is also present in the dust continuum emission, suggesting a common origin of both the dust and gas emission in this region. 
We focused on the overall structure of the emission and left a more detailed analysis of the gas kinematics to a future paper (Van der Wiel et al. in prep.). 

\section{Dust radiative transfer modeling}
\label{sect:rad_transf}
This Section describes the setup of the radiative transfer code \texttt{RADMC-3D}, including a description of the dust opacity, model grid and the envelope dust density model used. At the end of the Section, we present our investigation of the temperature structure difference between a single- and a binary protostar, where the envelope dust density model was used.
We constructed 3D dust and gas line radiative transfer models consisting of a large envelope with a radius of 8$\times10^{3}$~AU based on \mbox{\citet{2002A&A...390.1001S}}, two disks around the protostars, as well as a dust filament between the two protostars. 
The disk's dust densities were modeled with both the model protoplanetary disk (PP-disk) structure known from Class II sources \mbox{\citep[e.g.,][]{2012ApJ...744..162A}} and the density solution from the angular momentum conservation of an infalling, rotating collapse \mbox{\citep{1984ApJ...286..529T}}. This infalling, rotating collapse density structure is hereafter referred to as a rotating toroid model, based on the toroidal shape appearing at the centrifugal radius \citep{1984ApJ...286..529T}, which is the radius where the infalling material with the largest angular momentum impacts the midplane.  
In terms of luminosity, all models, whether they used one or two radiation sources, were limited to a total luminosity of 21~$L_{\sun}$. This luminosity is based on integration of the SED using data from \textit{Spitzer}, ISO-LWS, \textit{Herschel}/SPIRE and submillimeter data \citep{2005ApJ...632..371C, 2004A&A...418..607C, 2014PhDT.......304M, 2002A&A...390.1001S}, where the luminosity is estimated to be $21\pm5$~L$_{\sun}$ \mbox{\citep{2016A&A...595A.117J}}. This luminosity assumes a distance of 120~pc, based on recent distance estimates of the bulk of the $\rho$ Oph cloud from extinction measurements \mbox{\citep{1998A&A...338..897K, 2008A&A...480..785L}} and VLBI parallax \mbox{\citep{2008ApJ...675L..29L}}. 
The interbinary distance was fixed to 636~AU for all models, based on the angular distance of $\sim$5.3$^{\prime\prime}$ between the peaks of the 868~$\upmu$m ALMA image, with the caveat that this distance, in reality, is only the projected, that is, the minimum distance between the sources.

\subsection{\texttt{RADMC-3D}}
To determine the dust temperature and to produce continuum images, the radiative transfer code \texttt{RADMC-3D} \citep[][v. 0.40]{2012ascl.soft02015D}\footnote{\url{http://www.ita.uni-heidelberg.de/~dullemond/software/radmc-3d/}} was used. \texttt{RADMC-3D} uses the \mbox{\citet{2001ApJ...554..615B}} algorithm to perform a Monte Carlo simulation of photon propagation from one or more radiation sources, through a predefined grid of dust density cells, in order to estimate the dust equilibrium temperature. This equilibrium is a balance between the incoming radiative energy and the blackbody emission from the dust cell itself. The code setup in this work assumed that all the dust energy originated from the protostars through irradiation, and not from any internal energy source such as viscous accretion, or any external radiation source. The precision of the resulting dust equilibrium temperature depends, among other parameters, on the number of photons used in the radiation simulation (this differed between model setups, see Table \ref{tab:modelsetup}). 
We ignored scattering, since the absorption opacity dominates over the scattering opacity in the submillimeter regime \citep{2010ApJ...710..470D}. 
\texttt{RADMC-3D} was also used to calculate the SED (see Appendix \ref{sec:sed}) and to produce synthetic continuum images (see Section \ref{sec:disk_mods}).
\subsection{Grid}

\begin{table*}[!htbp]
  \begin{tiny}
    \caption{\texttt{RADMC-3D} model parameters}
    \centering
    \begin{tabular}[!htbp]{l l l l l}\toprule
    \midrule
    Parameter & Description & Value\\
    \midrule
    \textit{Grid}\\
    \midrule
    $r_{\mathrm{out}}$ & Envelope radius & 8$\times10^{3}$~AU  \\
    $r_{\mathrm{in}}$ & Grid start radius & 5~AU\\ 
    $r_{\mathrm{plat}}$ & Radius of the density plateau & 600~AU\\
    $n_{\mathrm{\uptheta}}$ & Number of grid cells in the polar range & 131\\
    $n_{\mathrm{\upphi}}$ & Number of grid cells in the azimuthal range & 131\\
    \midrule
    \textit{Temperature analysis}\\
    \midrule
    $n_{\mathrm{r,\:out}}$ & Number of grid cells in the outer radial region ($r$ $>$ 600 AU)  & 88 \\
    $n_{\mathrm{r,\:in}}$ & Number of grid cells in the inner radial region ($r$ $\leq$ 600 AU) & 110\\
    $n_{\mathrm{photons}}$ & Number of photons used in thermal Monte Carlo process & 5$\times10^{7}$ \\
    $L_\mathrm{A}$ & Luminosity of source~A & 3 -- 20 L$_{\sun}$ \\
    \textit{One radiation source} \\
    $r_{\mathrm{oct}}$ & Octree refinement radius around source & 400 AU \\
    $n_{\mathrm{octree\:levels}}$ & Octree refinement number & 1 \\
    \textit{Two radiation sources} \\
    $r_{\mathrm{oct}}$ & Octree refinement radius around each source & 250 AU \\
    $n_{\mathrm{octree\:levels}}$ & Octree refinement number & 2 \\
    \midrule
    \textit{IRAS~16293 model}\\
    \midrule
    $n_{\mathrm{r,\:out}}$ & Grid cells in the outer radial region ($r$ $>$ 600 AU) & 22 \\
    $n_{\mathrm{r,\:in}}$ & Grid cells in the inner radial region ($r$ $\leq$ 600 AU) & 110 \\
    $\rho_{0, \: \mathrm{env}}$ & Envelope reference density at $r_{0, \:\mathrm{env}}$ & 2.5$\times 10^{-14}$~g~cm$^{-3}$ \\
    $r_{0, \:\mathrm{env}}$ & Envelope reference radius & 1~AU\\
    $p_{\mathrm{env}}$ & Envelope density power-law exponent & 1.7\\
    $M_\mathrm{A}$ & Source A stellar mass & 1 M$_{\sun}$ \\
    $M_\mathrm{B}$ & Source B stellar mass & 0.1 M$_{\sun}$ \\
    $^{\dagger}T_{\mathrm{eff}}$ & Star surface temperature & 5000~K \\ 
    $L_\mathrm{A}$ & Source A luminosity & 3 -- 20 L$_{\sun}$\\ 
    $r_{0, \: \mathrm{disk}}$ & PP-disk reference radius & 10~AU \\
    $r_{\mathrm{inner\:disk}}$ & Disk inner radius & 1~AU\\
    $r_\mathrm{disk,A}$ & Disk A radius & 150~AU\\
    $r_\mathrm{disk,B}$ & Disk B radius & 50~AU\\
    \midrule
    \textit{Disk constraints}\\
    \midrule
    $F_{\mathrm{peak,\:A}}$ & Peak flux density toward source~A & 1.0 Jy beam$^{-1}$ \\
    $F_{\mathrm{peak,\:B}}$ & Peak flux density toward source~B & 2.0 Jy beam$^{-1}$ \\		
    Disk A aspect ratio & \dots & 1.64 \\		
    $N_{\mathrm{H_2,\:B}}$ & Disk B column density & $\geq$ 1.2$\times$ 10$^{25}$~cm$^{-2}$ \\
    \midrule 
    \textit{PP-disk models}\\
    \midrule
    $\Sigma_\mathrm{A}$ & Parameter space for reference surface density of PP-disk A at 10 AU & 0.1 -- 5.0 g~cm$^{-2}$ \\
    $\Sigma_\mathrm{B}$ & Parameter space for reference surface density of PP-disk B at 10 AU & 1.0 -- 12.0	g~cm$^{-2}$ \\
    $\Psi$ & Parameter space of the PP-disks flaring constants & 0 -- 0.25\\
  $p$ & Parameter space of the PP-disks dust density radial power-law exponents & 0.5 -- 1.5 \\
    $H_{0, \mathrm{A}}$ & Parameter space of PP-disk A scale-height at 10 AU & 0.85 -- 2.0~AU \\
    $H_{0, \mathrm{B}}$ & Parameter space of PP-disk B scale-height at 10 AU & 0.85 -- 3.0~AU \\
    $i_\mathrm{A}$ & Parameter space of PP-disk A inclination, 90$^{\circ}$ is edge-on & 30 -- 90$^{\circ}$ \\ 
    $n_{\text{lev,\:disk}}$ & Octree refinement level of entire disk & 2\\
    $n_{\text{lev,\:in\:disk}}$ & Octree refinement level in innermost region & 4 \\ 
    $n_{\mathrm{photons}}$ & Number of photons used in thermal Monte Carlo process & $10^{7}$ \\

    \midrule	
    \textit{Rotating toroid models} \\
    \midrule
    $r_{c,\:\mathrm{A}}$ & Parameter space of the centrifugal radius of rotating toroid A & 50 -- 100~AU \\
    $r_{c,\:\mathrm{B}}$ & Parameter space of the centrifugal radius of rotating toroid A & 5 -- 40~AU \\
    $\dot{M}_\mathrm{A}$ & Parameter space of the mass accretion rate into source~A & 1.0$\times 10^{-6}$ -- 5.5$\times 10^{-6}$ M$_\sun$ yr$^{-1}$ \\
    $\dot{M}_\mathrm{B}$ & Parameter space of the mass accretion rate into source~B & 1.0$\times 10^{-6}$ -- 5.5$\times 10^{-6}$ M$_\sun$ yr$^{-1}$ \\
    $n_{\text{lev,\:disk}}$ & Octree refinement level of entire disk & 2\\
    $n_{\text{lev,\:in\:disk}}$ & Octree refinement level in innermost region & 2 \\ 
    $n_{\mathrm{photons}}$ & Number of photons used in thermal Monte Carlo process & $10^{6}$ \\

    \midrule	
    \textit{Dust arc} \\
    \midrule
    $\rho_0$ & Reference density & 2.5 $\times 10^{-17}$ g~cm$^{-3}$ \\
    $r_{\text{out}}$ & Filament radius from filament center & 130~AU \\
    Filament semi-minor axis & \dots & 170~AU\\
    Filament semi-major axis & \dots & 300~AU\\
    \bottomrule
    \end{tabular}
    \tablefoot{Model parameters for each setup. $L_\mathrm{B}$ is always defined as $L_\mathrm{B}$ = 21.0 L$_\sun$ - $L_\mathrm{A}$. 
    $^{\dagger}$The radiation spectra of both source~A and B are treated as blackbodies with a temperature of 5000~K, following \citet{2002A&A...390.1001S}.}
    \label{tab:modelsetup}
\end{tiny}
\end{table*}
We constructed a spherical grid of cells for use in the \texttt{RADMC-3D} radiative transfer calculations, resolved linearly in the polar and azimuthal angle ranges, while the radial dimension was divided into an inner and outer region. The inner region with constant dust density (i.e., $r \leq r_{\mathrm{plat}}$, see Eq. \ref{eq:rhoplat}) was resolved linearly, while $r > r_{\mathrm{plat}}$ (containing the bulk mass of the cloud) was resolved logarithmically. In the disk components, of the global model, octree refinement was performed (splitting a grid cell into eight new grid cells), with two refinement levels, to ensure that the density gradients of the disk component models were resolved, see Table \ref{tab:modelsetup}. If a PP-disk model was used, then the inner 10~AU of the disk was refined two times further, for a total of four octree refinements, to resolve the inner gap and nearby structure properly, as geometric effects such as self-shadowing onto the outer disk became important.
\subsection{Dust opacity \& temperature}
\label{subsec:dust_opac_and_temp}
We used dust opacities from \mbox{\citet{1994A&A...291..943O}}, which are good approximations for dense cores \citep{1999ApJ...522..991V,2002ApJ...575..337S,2011ApJ...728..143S}.
Opacity tables of dust with thin ice-mantles and another with bare dust grains were used, assuming coagulated dust grains with an ambient gas number density of 10$^6$~cm$^{-3}$.
These dust opacity tables are generally used for models of dust around deeply embedded protostars, including IRAS~16293 \citep[e.g.,][]{2002A&A...390.1001S, 2005ApJ...631L..77J, 2010A&A...519A..65C}.
The dust opacity was allocated in the cells in a self-consistent manner, with the first \texttt{RADMC-3D} thermal Monte Carlo photon propagation done in a model of purely icy-dust opacities. Grid cells with temperatures above the assumed water sublimation temperature were then allocated bare-grain dust opacities for the next photon propagation. We did not change the sublimation temperature to account for the local environment in terms of pressure but instead used a single sublimation temperature of 90~K \citep{1993ApJ...417..815S}. 
As the cell opacities were redefined, the cell dust equilibrium temperatures after the next photon propagation will be different, with some dust cells shifting from below to above 90~K and vice versa, that is, an opacity allocation error. For both disk component model types, this error was found to be negligible after two recursions (5 and 0.1~\% relative to the total bare-grain dust mass, at $1 \sigma$, for the PP-disk model and the rotating toroid model, respectively). The higher error in the PP-disk model, even with ten times more photon packages than the rotating toroid model (Table \ref{tab:modelsetup}), came from the very high dust densities necessary to match the observed peak flux density, especially in disk B. In contrast, the relatively more uniform dust density distribution in a rotating toroid model ensured good photon statistics in \texttt{RADMC-3D}, with a much lower photon number.
\\
Therefore, two recursions of the bare-grain opacity allocation, that is, three Monte Carlo dust temperature estimations in total were performed in \texttt{RADMC-3D}, for a single parameter set. The dust temperature estimated in the final photon propagation was then used for all later computations with that parameter set.
\subsection{Dust envelope model}
The dust envelope was modeled with a 1D radial density power-law density distribution, as the envelope has been well-fitted previously by such models \mbox{\citep[e.g.,][]{2002A&A...390.1001S}}. More complicated circumbinary envelope structures, such as an infalling, rotating circumbinary envelope were not considered, as we focused on the innermost region of IRAS~16293 in this work.
Recent observations of GG Tau \mbox{\citep{2014Natur.514..600D}}, as well as simulations of the same system \mbox{\citep{2016ApJ...827...93N}} suggest that binary star formation results in partial clearing of the material in the innermost region where the protostellar binary resides. Based on the 868 $\upmu$m continuum observation, we do not expect the interior to be completely evacuated of dust between the disks, but depleted instead, due to accretion into the two protostars and dynamical effects, as described by \mbox{\citet{2016ApJ...827...93N}}. A dust density plateau was used to imitate the mass depletion inside a given radius $r_{\mathrm{plat}}$, meaning that we used a constant, rather than a radial power-law, density distribution in this inner region, while keeping the dust density distribution in the outer envelope as a power-law:
\begin{empheq}[left={\rho(r)=\empheqlbrace}]{alignat=2}
\begin{aligned}
\:&\rho_0 \left(\frac{r}{r_0}\right)^{-p} &, \:\text{if}\:r \geq r_{\mathrm{plat}} \\
\:&\rho_0 \left(\frac{r_{\mathrm{plat}}}{r_0}\right)^{-p} &, \:\text{if}\:r <  r_{\mathrm{plat}}.
\label{eq:rhoplat}
\end{aligned}
\end{empheq}

Where $\rho$ is dust density, $r$ is the distance to the model center, $\rho_0$ is the density at distance $r_0$ and $r_{\mathrm{plat}}$ is the radius of the density plateau. $r_{\mathrm{plat}}$ was fixed to 600~AU in all models, a distance we chose to ensure that the entire dust filament structure resided in this density plateau. We fixed $p$~=~1.7, taken from \cite{2002A&A...390.1001S}. 

\subsection{Envelope temperature structure}
\label{sec:temp_struct}
In order to compare the temperature structure between a single star and a binary system, we used the dust envelope model from Eq. \ref{eq:rhoplat} with one or two radiation sources in \texttt{RADMC-3D}, without including the filament and disk component models. 
We were interested in the volume of regions with dust above 30, 50 and 90~K, as these temperatures roughly correspond to the sublimation temperatures of CO, H$_2$CO and H$_2$O, respectively \citep{2015A&A...579A..23J, 2001A&A...372..998C, 2001MNRAS.327.1165F, 1993ApJ...417..815S}.

These gaseous species are often used as probes of the gas and dust in observations of star-forming regions. We followed \mbox{\citet{2015A&A...579A..23J}} and used the sublimation temperature of CO at 30~K, assuming that the CO is mixed with water-ice on the grain surface, which is more realistic in the protostellar environment than pure CO ice, which has a lower binding energy and thus a lower sublimation temperature of 20 K \mbox{\citep{2015A&A...579A..23J}}.

\begin{figure*}[ht]
  \centering
      \includegraphics[height = 0.6\textheight, width=1.0\textwidth, keepaspectratio]{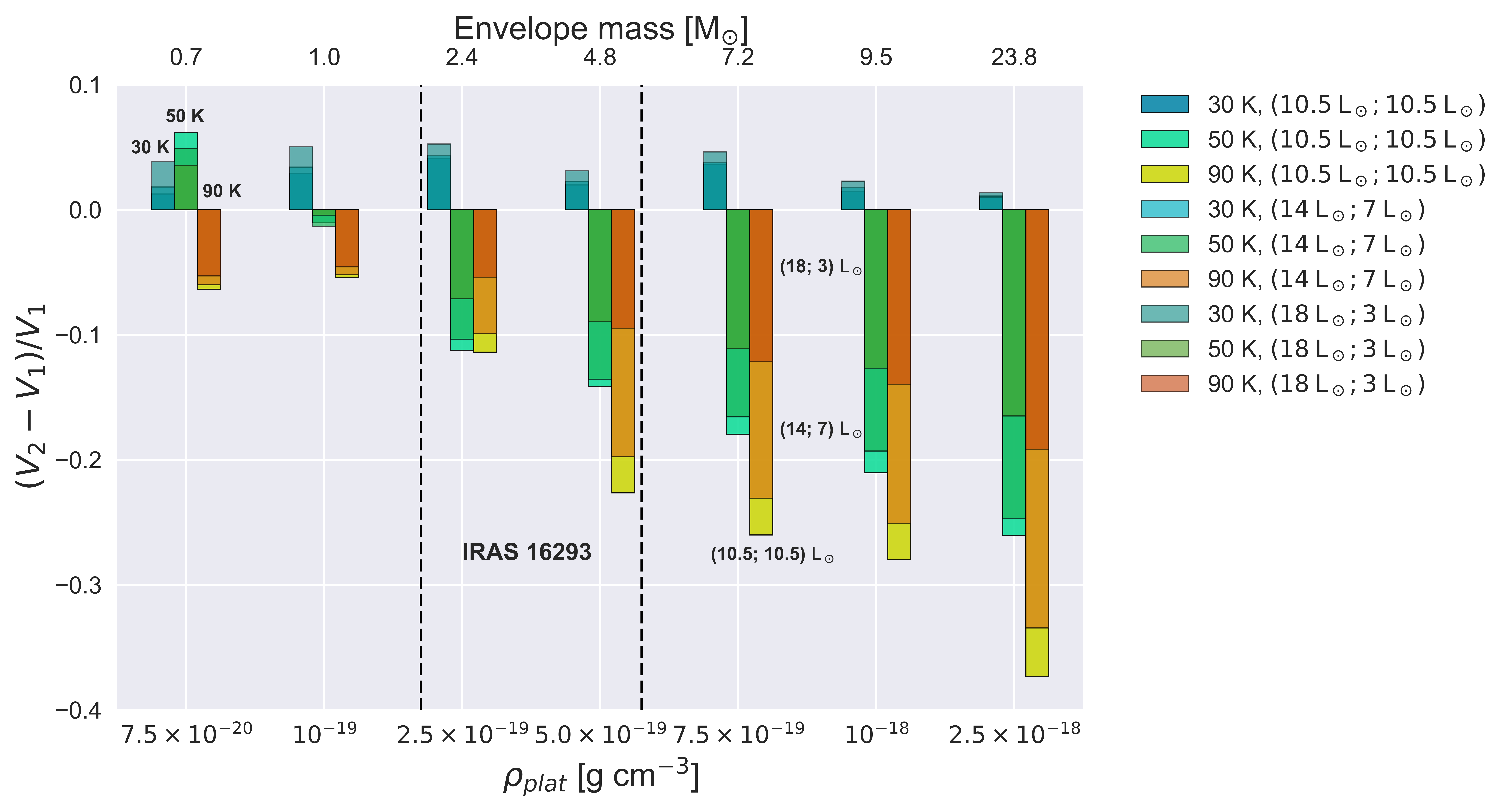}
      \caption{Relative differences in volumes of dust above different temperatures. The volume difference between the single star temperature volume, $V_1$, and binary star temperature volume, $V_2$, is shown, normalized to $V_1$, i.e. $\tau-0.2$ means that the total temperature volume is 20 \% lower if there are two stars compared to one. The envelope mass range is 0.7 -- 24 M$_{\sun}$, consistent with the dense core mass range observed in \mbox{\citep{2007A&A...462L..17A}}. The relevant mass range for IRAS~16293 is 2.4 -- 4.8~M$_{\sun}$, (our IRAS~16293 envelope model has a mass of $\sim$4~M$_\sun$). There is a clear tendency for a single star to result in more dust heated above $\sim$~50~K.}
      \label{fig:vol_norm}      
\end{figure*}

A walk through the parameter space of the dust reference density $\rho_0$, with $p = 1.7$, was performed with $r_{\mathrm{plat}}$ = 600~AU, $L_*$~=~21~L$_{\sun}$ (if one source) and $L_1$ = 10.5, 14.0 or 18.0 L$_{\sun}$, which means that, accordingly $L_2$ would be 10.5, 7.0 or 3.0 L$_{\sun}$, in the case of two sources.
The most important value is the central density plateau, $\rho_{\mathrm{plat}}$, as the 90~K region, almost all of the 50~K region and the bulk of the 30~K region resides here (Fig. \ref{fig:temp_invest} in Appendix \ref{app:CO}).
A given volume difference in two different $\rho_{\mathrm{plat}}$ scenarios corresponds to a higher absolute difference in the mass of dust above the given temperature threshold (Fig. \ref{fig:vol_norm}).

Across the sampled parameter space, as seen in Fig. \ref{fig:vol_norm}, the single star scenario has larger 90~K volumes than the binary and larger 50~K volumes for regions of higher density. The differences in 50~K go up to $\sim25~\%$, normalized to the volume of dust above 50~K for a single star. This is more pronounced for the 90~K volumes, where the relative difference is almost up to $40~\%$.
In the 30~K region, the binary system generally produced a slightly larger volume than the single star, with modest differences of 2 -- 5~\% in the total volume. In the case of IRAS~16293, with a total model mass of $\sim$4 M$_{\sun}$ in this work, the differences are on the order of 5 -- 20~\% in the 50 -- 90~K volumes.

Therefore, for observations of gaseous species such as CO, H$_2$CO and H$_2$O, where the abundances in the gas-phase are expected to increase with sublimation from the dust ice-mantles, two radiation sources should be included in the modeling of IRAS~16293, and of binaries with similar separations, as the total amount of sublimated gas will otherwise be overestimated.
If one star of the binary dominates the luminosity, the single-source approximation becomes more valid (Fig. \ref{fig:vol_norm}) and the overestimation of the sublimated gas becomes less important. The implications of these results for previous 1D chemical models of the hot corinos in IRAS~16293 are discussed further in Section \ref{sec:disc}.

\section{Dust density structures of IRAS~16293}
\label{sec:dustmodIRAS}
In this Section we present our analysis of the ALMA dust continuum image of IRAS~16293 at 868~$\upmu$m, where we investigated the presence of disks, or disk-like structures, and a dust filament, using a 3D dust density model.
The disk component models were fixed to 150 and 50~AU around source~A and B, respectively, based on visual inspection of the angular sizes and emission morphology of the dust continuum. The protostellar masses of source~A and B are unknown. 
\mbox{\citet{2004ApJ...617L..69B}} and \mbox{\citet{2011A&A...532A..23C}} estimate $M_{A}\sim$1~M$_{\sun}$, assuming the observed gas lines originate from infalling material, while the narrow linewidths observed by \mbox{\citet{2011A&A...532A..23C}} toward source~B provide an upper limit of 0.1~M$_\sun$, under the same assumption. 
\mbox{\citet{2012A&A...544L...7P}} resolve the region around source~A with a 2.2$^{\prime\prime}\times$ 1.0$^{\prime\prime}$ beam and interpret a Position-Velocity (PV) diagram of methyl formate (CH$_{3}$OCHO) and ketene (H$_2$CCO) as Keplerian rotation around a central object of 0.53~M$_{\sun}$ and a disk inclination close to edge-on, based on the linewidths (though the inclination is not constrained). \mbox{\citet{2016ApJ...824...88O}} use gas line observations toward source~A of higher spatial resolution (0.6$^{\prime\prime}\times0.5^{\prime\prime}$ beam) to conclude that a large part of the material around source~A is undergoing a rotating collapse, down to 40~--~60~AU, within which a Keplerian disk is speculated to lie. 
The inclination of the rotating collapse is not constrained, with best fits to their ballistic model between 30$^{\circ}$ and 70$^{\circ}$ (where 90$^{\circ}$ is edge-on). The derived mass range by \mbox{\citet{2016ApJ...824...88O}} of 0.5--1.0 M$_{\sun}$, is dependent on the inclination and centrifugal barrier radius. 
The mass of 0.53~M$_{\sun}$ reported by \mbox{\citet{2012A&A...544L...7P}} assumes exactly edge-on configuration and has lower spatial resolution than \mbox{\citet{2016ApJ...824...88O}}, thus arguably interpreting material undergoing a rotating collapse as part of a Keplerian disk. 
Also, the analysis of the velocity gradients on 50~--~400~AU scales around source~A by \mbox{\citet{2014ApJ...790...55F}}, shows a rotating structure which cannot be explained by simple Keplerian rotation around a point mass but needs to take the enclosed mass into account. We used $M_{A}$~=~1.0~M$_{\sun}$ in this work.
For source~B, \mbox{\citet{2012A&A...544L...7P}} find that the narrow linewidths from the gas in this region are consistent with a face-on disk. The aspect ratio upper limit of 1.1 from the disk emission around source~B \mbox{\citep{2016A&A...595A.117J}} is also consistent with a nearly face-on disk structure. Thus, the analysis by \mbox{\citet{2011A&A...532A..23C}} with a derived mass of $\leq$~0.1 M$_{\sun}$ seems valid and was used in this work. 
There have been reports of source~A itself being a binary, based on cm emission \mbox{\citep{1989ApJ...337..858W, 2010ApJ...712.1403P}} and submm emission \mbox{\citep{2005ApJ...632..371C}} with a projected distance of roughly 0.4$^{\prime\prime}$. The PILS program, with a comparable beam \mbox{\citep{2016A&A...595A.117J}}, does not find evidence of the split into separate submm sources reported by \mbox{\citet{2005ApJ...632..371C}}, but this could be due to optically thick dust emission, as the PILS continuum image (868 $\upmu$m) is at a shorter wavelength than the \mbox{\citet{2005ApJ...632..371C}} image ($\sim$1~mm). 
While it is unclear whether source~A is a binary or not, we modeled it as a single radiation source.

\subsection{Disk component model constraints}
The PILS interferometric beam of 0.5$^{\prime\prime} \times 0.5^{\prime\prime}$ is comparatively large compared to the continuum emission at 868 $\upmu$m around the two radiation sources, with the strongest continuum emission extending roughly 0.5$^{\prime\prime}$ around source~B and with a semi-major axis of 1$^{\prime\prime}$ for the elliptical emission around source~A (Fig. \ref{fig:dust_cont}). 
Since disk A appears close to edge-on, it should be possible to investigate the vertical distribution of dust with the continuum image. The model continuum emission around source~A was fitted with a 2D Gaussian in a 2$^{\prime\prime} \times 2^{\prime\prime}$ box centered on the peak flux density pixel on source~A, from which the aspect ratio of the semi-major axis to the semi-minor axis was derived from the fitted FWHM in the two dimensions. The observed 868 $\upmu$m continuum emission in a 2$^{\prime\prime} \times 2^{\prime\prime}$ box around source~A was found to have an aspect ratio of 1.64, which we required the model to match within 10~\% along with the peak central flux density of source~A within 10~\%. The box of 2$^{\prime\prime} \times 2^{\prime\prime}$ was chosen to avoid the observed warped structure around source~A at larger scales (likely belonging to the dust filament, see Fig. \ref{fig:dust_cont}), a disk morphology we did not attempt to recreate in this work. We only required our continuum modeling to match the peak flux density of source~B within 10~\%, as more detailed structures are hard to distinguish with our resolution.  
The column density limit toward the optically thick source~B peak continuum emission, which \mbox{\citet{2016A&A...595A.117J}} derive to be N$_{\text{H$_2$}}$~\textgreater~$\:1.2\times10^{25}$~cm$^{-2}$, was used as well.
Any model which satisfied these constraints also needed to reproduce the general features in both the dust continuum emission arc between the sources and the three CO isotopologue maps. Including these, six constraints were available per disk component model (the four emission maps, plus the two individual disk constraints).

\subsection{Dust filament}
\label{sec:dust_arc}
The arc of continuum emission (Fig. \ref{fig:dust_cont}) is assumed to arise from a local dust density enhancement. We modeled this dust filament as an elliptical tube, extending between and enveloping the sources. Around each source, the dust filament was extended spherically up to 130~AU. The density was modeled as a radially dependent power-law distribution of $\rho = \rho_0$($r$/AU)$^{-1/4}$ as measured from the tube center, where $\rho_0$ is the reference density at 1~AU. As such the filament dust mass density was made to be independent of the distance to either source.
The disk components were carved out inside the dust filament model, in a sphere of 150 and 50~AU around source~A and B, respectively. No attempts were made to satisfy boundary conditions, resulting in density jumps in the model filament-disk boundary regions.
For simplicity, the dust filament structure was kept in the plane of the sky, as were the two protostars. We fixed the filament model with a $\rho_0$ value, which matched the observed central arc emission with several different luminosity ratios (Fig. \ref{fig:cont_mod_vs_obs}), see Table \ref{tab:modelsetup}.
We performed a sanity check using the maximum observed emission of 0.056~Jy~Beam$^{-1}$ in the filament to calculate a column density based on Eq. 5 in \citet{2007ApJ...659..479J}, assuming a dust temperature of 30~K. The dust column density of the filament was derived to be 9.9~g~cm$^{-2}$, while the model column density was 9.7~g~cm$^{-2}$, assuming a gas-to-dust mass ratio of 100. This match is not surprising as we used the same opacities as \citet{2007ApJ...659..479J} and the optical depth through the model filament is low ($\sim$0.2).

\subsection{Disk component models}
\label{sec:disk_mods}
The disk components were modeled with both a PP-disk and a rotating toroid model.
In the PP-disk model, the dust distribution is based on hydrostatic equilibrium with the gravity of the host star \mbox{\citep[e.g.,][]{2012ApJ...744..162A}}.\\
The following equations all use cylindrical coordinates. The disk scale-height, $H_\mathrm{d}$, is
\begin{equation} 
  H_\mathrm{d}(r) = H_0(r/r_0)^{1+\Psi},
\end{equation}
where $r$ is the radius, $H_0$ is the scale-height at radius $r_0$ and $\Psi$ is the flaring constant, controlling the vertical distribution of the dust above the disk midplane. The disk surface density, $\Sigma$, is
\begin{equation}
\Sigma(r) = \Sigma_0(r/r_0)^{-p},
\end{equation}
where $p$ is a gradient parameter. Finally the dust density is 
\begin{equation}
  \rho_\mathrm{d}(r, z) = \frac{\Sigma(r)}{H_\mathrm{d}\sqrt{2\pi}}e^{-z^2/(2 H_\mathrm{d}^2)}, 
\label{eq:rho_complete}
\end{equation}
where $z$ is the height above the disk midplane.

The disk inner radius was fixed to 1~AU for both disks, with the dust density within 1~AU set to 10$^{-25}$~g~cm$^{-3}$ (n$_{\text{H}_2}$~$\sim$~3~cm$^{-3}$), effectively assuming a dust and gas evacuated inner gap. 
After the dust temperature of a given model was estimated in \texttt{RADMC-3D}, subsequent ray tracing produced a 868~$\upmu$m continuum image, which, after convolution with a 0.5$^{\prime\prime}\times0.5^{\prime\prime}$ beam using \texttt{MIRIAD}, was compared with the 868 $\upmu$m ALMA continuum image in Fig. \ref{fig:dust_cont}.
The disk orientation for source~B was fixed to face-on, while the inclination of PP-disk A was a free parameter.

Due to the high number of grid cells required to resolve the relevant regions and high photon numbers required for reliable dust temperature results, but most importantly the high densities in the inner disk regions of this work, the thermal Monte Carlo calculation execution time, on the available computing facilities, for a single model was very long, sometimes on the order of days. This problem was compounded by the iterative allocation of dust opacities, which requires three separate dust temperature estimations in \texttt{RADMC-3D}. Thus, standard model optimization approaches such as a Markov Chain Monte Carlo (MCMC) or a $\chi^2$ analysis of a large, uniform grid of model parameters were not feasible in this work and no parameter confidence intervals were derived. 
Instead, we identified parameter spaces not compatible with the current constraints and derived conclusions from these parameter space exclusions. 

\subsection{Model analysis}
\label{subsec:dust_res}
Starting with the PP-disk model, large steps were taken in the parameter space of $\Sigma_0$ and $H_0$ for the disk around each source, and inclination for disk~A (Table~\ref{tab:modelsetup}), while $p$ and $\Psi$ were fixed to 0.5 and 0.25, respectively (see Appendix \ref{app:p_psi_par} for more details). 
The final number of free parameters in the PP-disk model was four for disk~A ($\Sigma$, $H_0$, $i$ and source~A luminosity, $L_\mathrm{A}$) and two for disk~B ($\Sigma$ and $H_0$), since the source~B luminosity, $L_\mathrm{B}$, was determined by $L_\mathrm{A}$.

When only using the continuum image, the PP-disk model did not offer any meaningful constraints on disk~A and B. Due to the degeneracy between $H_0$ and $i$, only models with inclinations below 55$^{\circ}$ could be excluded. 
The peak flux values were degenerate with different $\Sigma_0$, $H_0$, $L_*$ and $i$ values and the column density toward source~B did not offer any constraints either, as almost all of our investigated parameter space satisfied the $N_{\text{H}_2}$ column density condition. 
\begin{figure*}[ht]
  \centering
      \includegraphics[height = 0.66\textheight, width=0.99\textwidth, keepaspectratio]{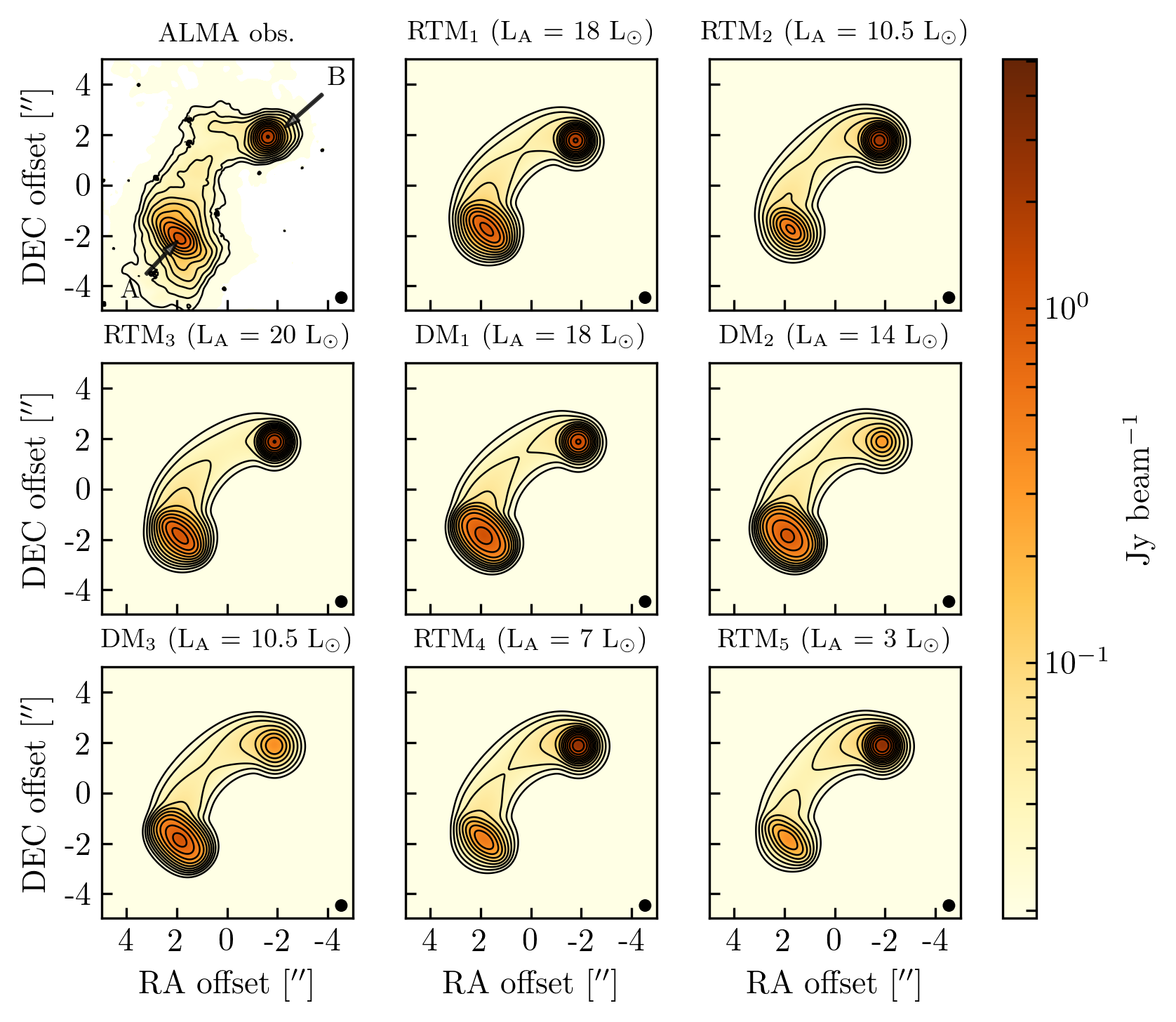}      
      \caption{Observed and synthetic dust continuum emission at 868 $\upmu$m for the different physical models summarized in \ref{tab:dust_mods}. Image is in log-scale and the contours are logarithmically divided between 0.02 and 2~Jy~beam$^{-1}$. DM denotes the PP-disk model, while RTM denotes the rotating toroid model.}
      \label{fig:cont_mod_vs_obs}
\end{figure*}
\begin{table*}
    \centering
    \caption{Example dust model parameters.}
    \begin{tabular}{l*{8}{c}r}\toprule
    \midrule
    \textit{PP-disk model}
    & $L_\mathrm{A}$ [L$_{\sun}$] & $L_\mathrm{B}$ [L$_{\sun}$] & $\Sigma_\mathrm{0,A}$ [g~cm$^{-2}$] & $\Sigma_\mathrm{0,B}$ [g~cm$^{-2}$] & $H_\mathrm{0,A}$ [AU]  & $H_\mathrm{0,B}$ [AU] & $i_\mathrm{A}$ [$^{\circ}$]\\    
\midrule
    DM 1 & 18.0 & 3.0 & 1.0 & 3.0 & 1.25 & 1.5 & 55\\
    DM 2 & 14.0 & 7.0 & 1.0 & 0.05 & 0.85 & 2.0 & 50 \\
    DM 3 & 10.5 & 10.5 & 1.0 & 0.05 & 1.25 & 2.0 & 60 \\
    \midrule
    \textit{Rotating toroid model}
    & $L_\mathrm{A}$ [L$_{\sun}$] & $L_\mathrm{B}$ [L$_{\sun}$] & $\dot{M}_\mathrm{A}$ [$M_{\sun}$ yr$^{-1}$] & $\dot{M}_\mathrm{B}$ [$M_{\sun}$ yr$^{-1}$] & $r_\mathrm{c,A}$ [AU] & $r_\mathrm{c,B}$ [AU]  \\
    \midrule
    RTM 1 & 18.0 & 3.0 & 3.75$\times 10^{-6}$ & $2.75\times 10^{-6}$ & 75 & 20 & & &\\
    RTM 2 &10.5 & 10.5 & 2.1$\times 10^{-6}$ & $2.1\times 10^{-6}$ & 50 & 20 & & &\\
    RTM 3 & 20.0 & 1.0 & 3.8$\times 10^{-6}$ & $4.6\times 10^{-6}$ & 75 & 20 & & &\\
    RTM 4 & 7.0 & 14.0 & 3.0$\times 10^{-6}$ & $2.0\times 10^{-6}$ & 75 & 20 & & &\\
    RTM 5 & 3.0 & 18.0 & 3.0$\times 10^{-6}$ & $2.0\times 10^{-6}$ & 75 & 20 & & &\\
    \bottomrule
    \end{tabular}
    \label{tab:dust_mods}
\end{table*}
Due to previous observations of infall toward both sources \citep{2011A&A...532A..23C,  2005ApJ...632..371C, 2012A&A...544L...7P, 2013ApJ...764L..14Z, 2016ApJ...824...88O}, we considered the rotating toroid model.  
The dust density distribution of this model can be found semi-analytically by defining discrete streamlines from the rotating envelope, which spiral inwards and impact opposing streamlines in the midplane, creating a disk structure. From \mbox{\citet{2009apsf.book.....H}},
\begin{equation}
  \rho(r,\theta)=\frac{\dot{M}\left(1+\frac{\mathrm{cos}\theta}{\mathrm{cos}\theta_0}\right)^{-1/2}}{4\pi\sqrt{GM_*r^3}}
\left(\frac{\mathrm{cos}\theta}{\mathrm{cos}\theta_0}+\frac{2\mathrm{cos}^2\theta_0}{r/r_\mathrm{c}}\right)^{-1},
\label{eq:rot_coll}
\end{equation} 
where $\dot{M}$ is the mass accretion rate onto the central object, $\theta_0$ is the starting polar angle of the streamline in question and $r_\mathrm{c}$ is the centrifugal radius. The density in each grid cell in our model was found by predefining $n_{\mathrm{stream}}$ streamlines with discrete $\theta_0$ values and evaluating them at $n_{\mathrm{r}}$ radial points, along the streamline. A nearest-neighbor search was performed, with the dust density of the \texttt{RADMC-3D} grid cell defined to be that of the nearest defined streamline point.
The rotating toroid model is effectively a 2D density model and a 3D velocity model \mbox{\citep{2009apsf.book.....H}}.

The rotating toroid model has a radially extended dust distribution with $\rho$ $\propto$ $r^{-0.5}$ and a much more vertically extended dust distribution than the exponentially decreasing vertical distribution of the PP-disk model (Eq. \ref{eq:rho_complete}). Consequently, the rotating toroid model has a much lower optical depth through the midplane, hence higher dust temperatures, than a PP-disk model with, for example, $H_{0}$~=~0.85~AU \citep[this example value was taken from the range in][]{2010ApJ...723.1241A}.

The centrifugal radius of the collapse around source~B ($r_\mathrm{c,B}$) was a free parameter between 5~--~25~AU, to allow a compact toroid within the defined model disk~B regime of 50~AU, fixed to face-on inclination. We fixed the inclination of the rotating toroid model around source~A to edge-on and used $r_\mathrm{c,A}$~=~75~AU, as it matched the aspect ratio at this inclination.
The free parameters of the source~A rotating toroid model were $\dot{M}_\mathrm{A}$ and $L_\mathrm{A}$, as $r_\mathrm{c,A}$ and $M_\mathrm{A}$ were fixed (Table \ref{tab:modelsetup}), while the source~B rotating toroid model had $\dot{M}_\mathrm{B}$ and $r_\mathrm{c,B}$ as free parameters (since $L_\mathrm{B}$ was determined by $L_\mathrm{A}$). We found that the observed peak flux density toward source~A and B could not constrain the luminosity or the mass accretion rate of either source, as these parameters are degenerate. The minimum column density toward source~B did not offer any constraint.
In conclusion, the rotating toroid model can satisfy the disk constraints, but cannot constrain $L_\mathrm{A}$ or $L_\mathrm{B}$, using continuum data alone (Table \ref{tab:modelsetup}).

\section{Gas line emission modeling}
\label{sec:linemod}
In order to constrain the nature of the disk-like emission further, and to obtain more constraints on the source luminosities, the $J$=3--2 line emission from $^{13}$CO, C$^{17}$O and C$^{18}$O of select dust models were modeled in \texttt{LIME} \mbox{\citep{2010A&A...523A..25B}}, a line radiative transfer code.
The dust temperature output from \texttt{RADMC-3D} was imported to \texttt{LIME} (Appendix \ref{app:radlime}) and by using the standard gas-to-dust mass ratio of 100 \mbox{\citep{1978ApJ...224..132B}} together with the given dust density model, the molecular hydrogen number density $n_{\text{H$_\text{2}$}}$ was found.
The gas line emission depends on the number density of the molecule undergoing a rotational transition, the CO isotopologue in our case, the number density of the other molecules in the gas which excite the molecule collisionally and the kinetic temperature which changes the collisional rates \mbox{\citep{2010A&A...523A..25B}} and thus the excitation of the CO isotopologue. In \texttt{LIME}, the number density of the colliders is approximated as $n_{\mathrm{H_2}}$, as H$_2$ is by far the most abundant species in the dense molecular cores where star formation takes place \mbox{\citep[e.g.,][]{1989ARA&A..27...41G}}. \texttt{LIME} performs non-LTE (Local Thermodynamic Equilibrium) line radiative transfer modeling, where the rotational level populations are found iteratively.
The CO isotopologue data files containing collisional rate coefficients \citep{2010ApJ...718.1062Y} were downloaded from the LAMDA\footnote{\url{http://home.strw.leidenuniv.nl/~moldata/}} database \mbox{\citep{2005A&A...432..369S}}. The collisional rate coefficients of the CO isotopologues with H$_2$ in these files include collisions with both ortho- and para-H$_2$.

In this work, we assumed that the dust and gas temperatures are coupled, which is usually a good approximation in the dense inner environments of the envelope \mbox{\citep{2002A&A...390.1001S}}, that were modeled here. 
The radius of the \texttt{LIME} model was set to $2\times10^{3}$~AU, as more distant, cold regions in the envelope will not contribute much emission due to freeze-out of CO. 

Optimally, the absorption features around each source and the major features of the interbinary CO isotopologue emission seen in Fig. \ref{fig:co_obs} were recreated in the synthetic moment zero maps, which would imply that our dust density model is consistent with the observed gas line emission. We did not attempt to match the exact absorption level as it depends not only on the molecule abundance but also on the gas-to-dust mass ratio and dust opacity, neither of which were constrained in this work. 

In the case of a PP-disk model, the velocity in the disk around each source was modeled as Keplerian rotation. Here the central stellar mass becomes important if there is any disk inclination toward the observer, as radial velocities cause spectral broadening, reducing the line optical depth. This, in turn, will result in higher integrated emission if there is any line self-absorption, which we expect around each source, based on the observations seen in Fig. \ref{fig:co_obs}. In the case of a rotating toroid model, the 3D velocity structure from \mbox{\citet{1984ApJ...286..529T}} was used.
For all regions outside the disk components, the velocity was modeled as simple free-fall,
\begin{equation}
v(r) = \sqrt{G M_{\text{tot}}/r},
\end{equation}
where $r$ is the distance to the model center, $G$ is the gravitational constant and $M_{\text{tot}}$ is the combined mass of source~A and B.
This approximation is valid at kAU scales but breaks down in the innermost region. The dust in the interbinary region should arguably be drifting toward either source, in accordance with the local gravitational field. 
Since only one dust opacity model can be set in \texttt{LIME} (v. 1.5), bare-grain dust opacities were used in the entire model, as the dust grains in the disk regions are mostly above 90~K in the rotating toroid models, and in the PP-disk model atmosphere. If we had used icy-grain opacities in the \texttt{LIME} models instead, the reduced absorption from the dust would result in increased flux density globally in the integrated gas emission maps, with the largest difference in the disk regions.

\begin{table*}
    \caption{\texttt{LIME} model parameters} 
    \centering
    \begin{tabular}{l l l l l}
    \hline
    \midrule
    Parameter & Description & Value \\
    \midrule
    Grid outer radius & \dots & $2\times10^3$~AU \\
    Grid start radius & \dots & 30~AU \\ 
    n$_{\mathrm{nodes}}$ & Number of grid nodes in the \texttt{LIME} calculations & $7\times10^4$\\
    CO/H$_2$ & Standard ISM abundance of CO & 10$^{-4}$ \\
    $^{13}$CO/H$_2$ & Standard ISM abundance of $^{13}\mathrm{CO}$ & 1.30$\times$10$^{-6}$ \\ 
    C$^{18}$O/H$_2$ & Standard ISM abundance of $\mathrm{C}^{18}\mathrm{O}$ & 1.79$\times$10$^{-7}$ \\ 
    C$^{17}$O/H$_2$ & Standard ISM abundance of $\mathrm{C}^{17}\mathrm{O}$ & 5.58$\times$10$^{-8}$ \\ 
    Freeze out factor & \dots & 0.01 \\ 
    Gas-to-dust-mass ratio & Used for all conversions from dust to gas mass & 100 \\
    \midrule
    CO isotopologues abundances of all models \\
    $X_{^{13}\mathrm{CO}}$ & $^{13}\mathrm{CO}$ abundance & 2.6$\times$10$^{-7}$ \\
    $X_{\mathrm{C}^{18}\mathrm{O}}$ & $\mathrm{C}^{18}\mathrm{O}$ abundance & 3.58$\times$10$^{-8}$  \\
    $X_{\mathrm{C}^{17}\mathrm{O}}$ & $\mathrm{C}^{17}\mathrm{O}$ abundance & 1.12$\times$10$^{-8}$ \\
    \bottomrule
    \end{tabular}
    \tablefoot{Given abundances assume that C and O atoms maximally combine into CO.}
    \label{tab:lime}
\end{table*}

\subsection{Abundance modeling}
\label{sec:abun}
It is assumed that the bulk of the CO isotopologue emission originates from molecules sublimated from the dust grains, represented with an abundance jump model in this work, similar to some of the previous chemical modeling of IRAS~16293 \mbox{\citep{2002A&A...390.1001S, 2004A&A...418..185S,2012A&A...539A.132C}}.
The individual CO isotopologue abundances relative to H$_2$, the main collision partner in our modeling, were found by using the C and O isotopologue abundances given by \mbox{\citet{1994ARA&A..32..191W}} (Table \ref{tab:lime}).
Standard abundance values from \mbox{\citet{1994ARA&A..32..191W}} were applied to regions with $T_{\mathrm{dust}}~\geq~30$~K where the CO is sublimated from the dust grains, and a drop factor of 10$^{-2}$ was used in regions with $T$~$<$~30~K, to imitate freeze-out. Examining the temperature complexity of a typical rotating toroid model (Fig. \ref{fig:two_sourc_disks_torus} in Appendix \ref{app:CO}), when disks and the dust filament are added to the model, we can see the expected emission shape from the CO isotopologues in Fig. \ref{fig:co_obs_mod}, based on the $T$~=~30~K~contour. 

\subsection{Modeled CO isotopologue maps}
\label{sec:lime_mods_fit}
\begin{figure*}[ht]
  \centering
      \includegraphics[height = 1.0\textheight, width=1.0\textwidth, keepaspectratio]{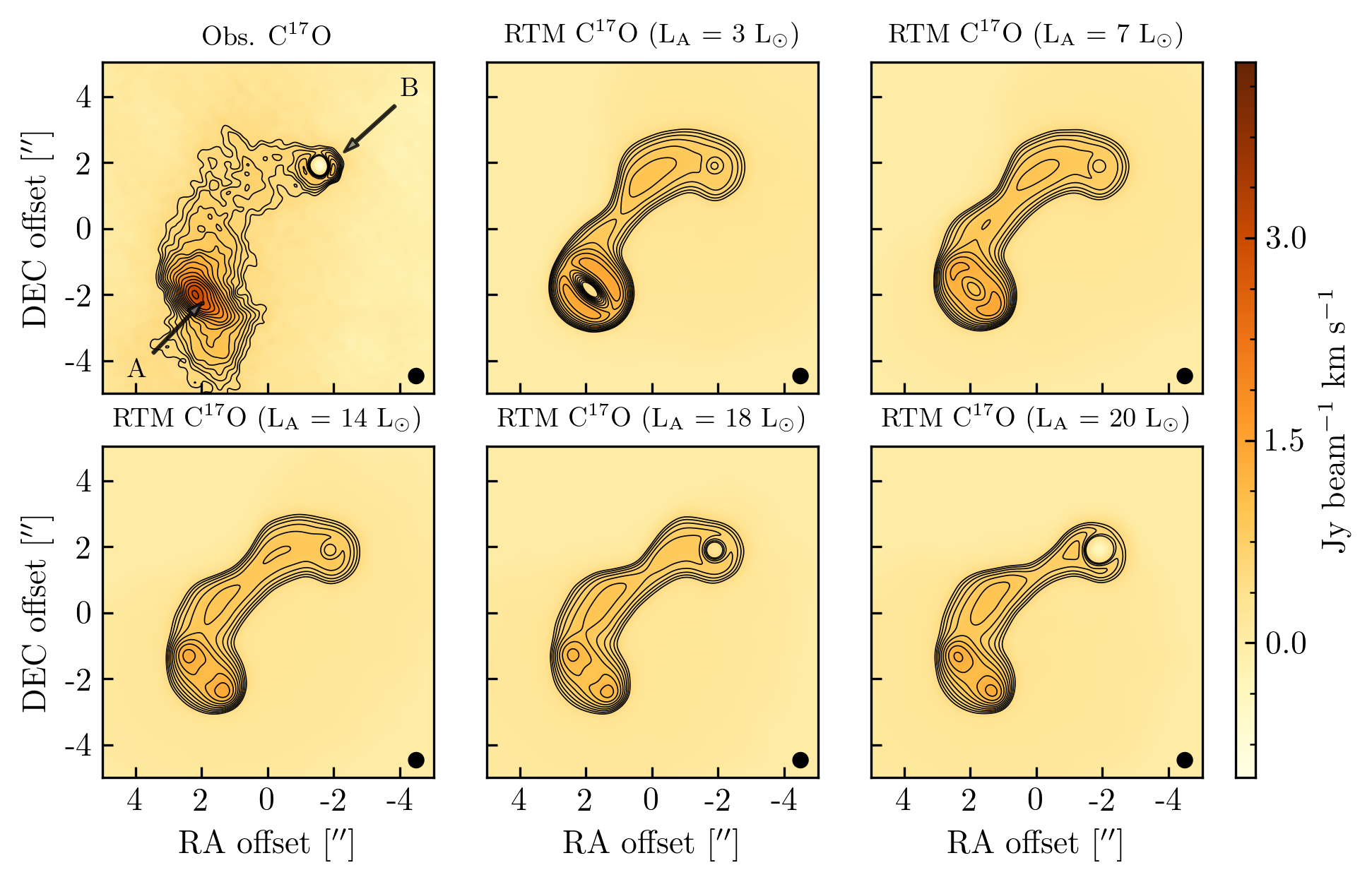}
      \caption{Observed and modeled C$^{17}$O zeroth moment maps, contour levels are divided logarithmically from 0.5 to 7.3 Jy~beam$^{-1}$~km~s$^{-1}$. All models are rotating toroid models matching the dust continuum peak flux densities for disk A and B.}
      \label{fig:CO_diff_lum}      
\end{figure*}

To constrain the luminosity ratio of the two protostars, we modeled the rotating toroid model at different luminosity values (Table \ref{tab:C17O_mods}), with parameter sets that satisfied the dust continuum constraints. The resulting synthetic CO isotopologue maps (Fig. \ref{fig:CO_diff_lum}) indicate that $L_\mathrm{A}$ $\geq$~18~L$_{\sun}$ in order to reproduce the observed tendency of concentrated emission toward source~A. 
We can deduce the luminosity dominance of source~A from the observations alone, as the truncation in the observed C$^{17}$O and C$^{18}$O maps occurs closer to source~B than source~A, which requires source~A to be more luminous, if we assume this truncation is due to the dust grain temperature dropping below the sublimation temperature of CO.
$L_\mathrm{A}$ $\geq$~18~L$_{\sun}$ is also consistent with the dust continuum model results (Fig. \ref{fig:cont_mod_vs_obs}).
Alternatively, the increase in continuum and gas line emission toward source~A could be due to geometric effects, such as an inclination of the stellar plane so source~A is closer to us than source~B, or simply that more dust material exists in the filament toward source~A. However, an increase in dust density would require $L_\mathrm{A}$ to be lowered, in order to match the observed continuum emission as well as increasing the optical depth. Both instances lower the gas temperature in the region near source~A, which creates a truncated region wherein CO is sublimated closer to source~A than source~B. 
In this sense, introducing more material in our filament model toward source~A should require $L_\mathrm{A}$ to be even higher than we currently find in order to match the observed CO isotopologue emission.
The inferred low luminosity of source~B implies that disk~B is denser than disk~A in order to dominate the dust submillimeter emission, which fits well with the observed absorption in the CO isotopologue emission around source~B, compared to the vicinity of source~A. 
\\
A representative rotating toroid model which satisfied the dust continuum constraints with $L_\mathrm{A}$~=~18~$L_{\sun}$ was modeled in \texttt{LIME}, together with a representative of the best matching PP-disk model at $L_\mathrm{A}$~=~14~$L_{\sun}$ (since $L_\mathrm{B}$~<~7~$L_{\odot}$ could not reproduce the observed peak flux density in our $H_0$ parameter space, though this does not rule out $L_\mathrm{B}$ < 7~$L_{\odot}$ by itself, see Fig. \ref{fig:disk_B_scale} in Appendix \ref{app:CO}) can be seen in Fig. \ref{fig:co_obs_mod}.
The first attempt with the standard abundances described in Table \ref{tab:lime}, resulted in significant absorption in a region toward source~B as expected, but also with some unexpected absorption toward source~A.
For the rotating toroid model, the C$^{18}$O model emission in the disk~A component has evident self-absorption throughout the midplane. 
After lowering the abundances by a factor of five, the model C$^{18}$O emission structure was more reminiscent of the observations, while C$^{17}$O did not reproduce the observed peaked emission in the disk~A component at any of the attempted abundances (Fig. \ref{fig:co_obs_mod}).

\begin{figure*}[ht]
    \centering
    \includegraphics[width=18cm, height=18cm, keepaspectratio]{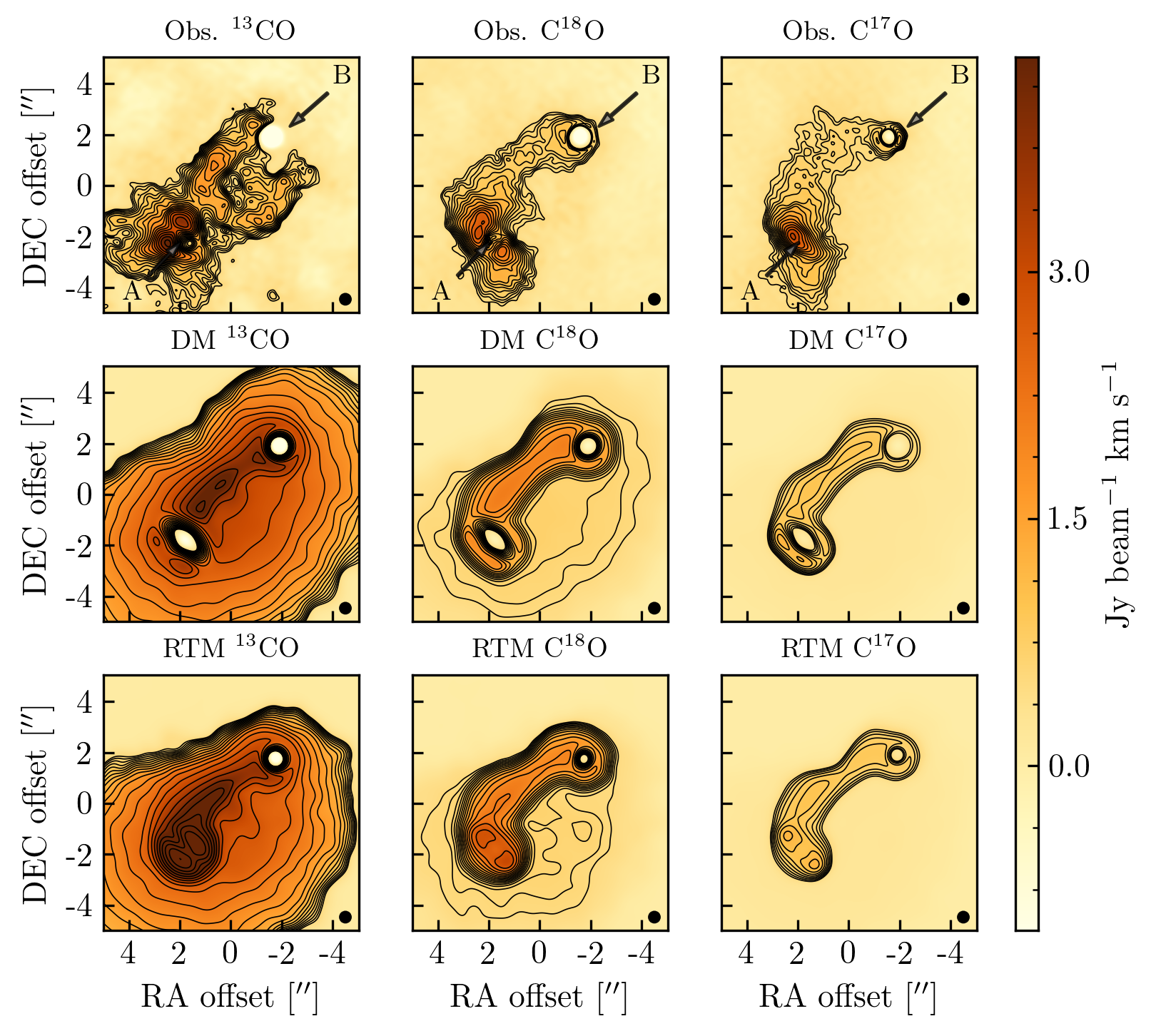}      
    \caption{Zeroth moment maps of the observed and synthetic CO isotopologue gas line emission. Contour levels are divided logarithmically from 0.5 to 7.3 Jy~beam$^{-1}$~km~s$^{-1}$. The RA and DEC offsets are relative to the phase center of the observations. See Table \ref{tab:lime_mods} for abundances. DM denotes the PP-disk model and RTM denotes the rotating toroid model.}
    \label{fig:co_obs_mod}
\end{figure*}

\begin{table*}
    \caption{\texttt{LIME} model parameters of Fig. \ref{fig:co_obs_mod}.} 
    \centering
    \begin{tabular}{l*{8}{c}r}\toprule
    \midrule
    Parameter & Description & Value \\
    \midrule
    \textit{Rotating toroid model} \\
    $L_\mathrm{A}$ & Luminosity of source~A & 18 L$_{\sun}$ \\
    $L_\mathrm{B}$ & Luminosity of source~B & 3 L$_{\sun}$ \\
    $\dot{M}_\mathrm{A}$ & Mass accretion rate & $3.75\times10{^{-6}}$ M$_{\sun}$ yr$^{-1}$ \\
    $\dot{M}_\mathrm{B}$ & Mass accretion rate & $2.75\times10{^{-6}}$ M$_{\sun}$ yr$^{-1}$ \\
    $r_\mathrm{c,A}$ & Centrifugal radius & 75 AU \\ 
    $r_\mathrm{c,B}$ & Centrifugal radius & 20 AU \\ 
    \midrule
    \textit{PP-disk model} \\
    $L_\mathrm{A}$ & Luminosity of source~A & 14 L$_{\sun}$ \\
    $L_\mathrm{B}$ & Luminosity of source~B & 7 L$_{\sun}$ \\
    $\Sigma_\mathrm{0,A}$ & Dust surface density at 10 AU & 0.6 g cm$^{-2}$ \\
    $\Sigma_\mathrm{0,B}$ & Dust surface density at 10 AU & 5.0 g cm$^{-2}$ \\
    $H_\mathrm{0,A}$ & Scale-height at 10 AU & 1.5 AU \\
    $H_\mathrm{0,B}$ & Scale-height at 10 AU & 2.0 AU \\    
    $i_\mathrm{A}$ & Disk A inclination & 60$^{\circ}$ \\
    \bottomrule
    \end{tabular}
    \label{tab:lime_mods}
\end{table*}
\begin{table}
    \caption{\texttt{LIME} model parameters of Fig. \ref{fig:CO_diff_lum}.} 
    \centering
    \begin{tabular}{l*{8}{c}r}\toprule
    \midrule
    Model & $\dot{M}_\mathrm{A}$ [M$_{\sun}$ yr$^{-1}$] & $\dot{M}_\mathrm{B}$ [M$_{\sun}$ yr$^{-1}$] \\ 
    \midrule
    $L_\mathrm{A}$ = 3 L$_{\sun}$ & 9.0$\times10^{-6}$ & 1.3$\times10^{-6}$ \\
    $L_\mathrm{A}$ = 7 L$_{\sun}$ & 5.5$\times10^{-6}$ & 1.4$\times10^{-6}$   \\
    $L_\mathrm{A}$ = 10.5 L$_{\sun}$ & 4.6$\times10^{-6}$ & 1.6$\times10^{-6}$   \\
    $L_\mathrm{A}$ = 14 L$_{\sun}$ & 4.2$\times10^{-6}$ & 1.85$\times10^{-6}$   \\
    $L_\mathrm{A}$ = 18 L$_{\sun}$ & 3.75$\times10^{-6}$ & 2.75$\times10^{-6}$   \\
    $L_\mathrm{A}$ = 20 L$_{\sun}$ & 3.8$\times10^{-6}$ & 4.6$\times10^{-6}$   \\
    \bottomrule
    \end{tabular}
    \tablefoot{All models are rotating toroids, using $r_\mathrm{c,A}$ = 75 AU and $r_\mathrm{c,B}$ = 20 AU. }
    \label{tab:C17O_mods}
\end{table}

The model emission qualitatively matches the observed $^{13}$CO bimodal emission structure above and below the disk midplane, 
as well as the two observed C$^{18}$O emission peaks near the corners of the disk~A component, but with less absorption in C$^{18}$O toward source~A than observed.
The rotating toroid model cannot, however, reproduce the C$^{17}$O centrally concentrated emission structure, but instead results in a bimodal distribution as seen in C$^{18}$O, near $r_\mathrm{c,A}$ ($r_\mathrm{disk}$~=~75~AU).
The PP-disk model fails to produce the emission peaks around source~A, in all isotopologues, as the PP-disk A region heavily absorbs the line emission, even when the standard abundances are lowered by a factor of five, despite the fact that we used the PP-disk model parameters with the lowest optical depth while matching continuum constraints.

When the PP-disk model standard abundances were lowered by a factor of ten or 15 instead (Fig. \ref{fig:class_ab_divs} in Appendix \ref{app:CO}) the C$^{17}$O emission from the filament became much weaker than observed, while the disk A region remained in absorption. The model absorption is then likely not due to line self-absorption, but rather originates from the optically thick dust. This suggests that the disk-like emission around source~A is composed of a relatively diffuse, vertically extended structure.

However, if larger amounts of CO are in front of disk~A, on similar spatial scales, then it could explain the extra line emission in the observations compared to the PP-disk model emission. Since the continuum emission is consistent with a PP-disk model, we cannot exclude the possibility that the disk-like emission structure around source~A could be from a flat disk, based on our model and observations alone. It is also possible that the presence of material in front of disk A could allow the rotating toroid models to use the standard ISM CO abundance and still qualitatively match the observations.

\section{Discussion}
\label{sec:disc}
The combined constraints from the dust continuum and CO isotopologue emission, necessitate very high PP-disk model scale-heights around source~B, $H_{0}$~>~3.0~AU at 10~AU, which is suggestive of a nonsettled disk with a large vertical distribution, meaning that it is indicative of a structure similar to that of the diffuse, rotating toroid model. These scale-heights are unprecedented for Class II disks, but similar to the one reported by \mbox{\citet{2013ApJ...771...48T}}, who find a best-fit PP-disk model to the Class~0 disk around L1527 with a scale-height of 48~AU at 100~AU. In comparison, when converting $H_0$ to the scale-height at 100 AU, the PP-disk model around source~B needs a scale-height of more than 53~AU at 100~AU, when $L_\mathrm{B}$~=~3~L$_{\sun}$, in order to reach the peak flux density value (Fig. \ref{fig:disk_B_scale} in Appendix \ref{app:CO}).
This value is significantly higher than the scale-heights at 100 AU of Class~I/II sources, which typically are 3~--~20~AU \mbox{\citep{2008ApJ...674L.101W, 2009ApJ...700.1502A, 2010ApJ...723.1241A}}. \mbox{\citet{2013ApJ...771...48T}} propose that the large scale-height of their best fitting PP-disk model of L1527 is approximating material from the envelope falling onto the disk and that this unusually vertically extended structure could reflect a stage before effective dust settling.
The disk-like structure around source~B could be similar to the L1527 disk in this regard, as infall toward source~B has previously been detected.

The CO isotopologue moment maps are not entirely qualitatively matched by a rotating toroid, but the fact that the standard CO isotopologue abundances needed to be lowered by a factor of five to avoid self-absorption in C$^{18}$O is interesting. \mbox{\citet{2016A&A...591A...3A}} use abundances an order of magnitude lower than the standard CO isotopologue abundances in order to match their observations of regions interior to the CO snowline in a survey of nearby Class~0 protostars.
In the vicinity of protostars and accretion shocks, the radiation field and chemistry is markedly different from the ISM, so deviations from the standard ISM abundances in the inner regions of a disk are not surprising \mbox{\citep[e.g., see][]{2009A&A...503..323V}}. However, because we base this necessity for a lower abundance on a single species, and since the self-absorption might simply reflect that the model is a poor approximation, a more thorough, quantitative investigation would be needed before making any firm conclusions on the CO abundance in the inner regions of IRAS 16293.\\

It has been suggested that a fraction of the CO is desorbed together with H$_2$O into the gas-phase at 90~K \mbox{\citep{2016A&A...591A...3A}}, due to trapping in the water-ice, which would allow higher abundances in the gas model for regions with T$_\mathrm{dust}$ $\geq$ 90~K than regions with 30~K $\leq$ T$_\mathrm{dust}$ $<$ 90~K. This should lead to an emission structure closer to that of the observed, with a more abrupt peak toward source~A and should be investigated in future modeling efforts.

Considering the unusual scale-heights of the PP-disk model, necessary to produce the peak flux density around source~B, a rotating toroid model is perhaps more representative of the actual dust distribution around source~B than a PP-disk model, as it better approximates the infalling material, unambiguously detected toward source~B \citep{2005ApJ...632..371C, 2012A&A...544L...7P, 2013ApJ...764L..14Z} as well as shocks \mbox{\citep{2011A&A...534A.100J}} which would be expected with a rotating collapse. 
It is also possible that both disks~A and B are transition structures, between an early rotating toroid structure and a later Keplerian disk, perhaps similar to the structure suggested by \mbox{\citet{2016ApJ...824...88O}}. 

In terms of the impact of the binarity of IRAS~16293 on earlier works using a single radiation source, we can consider \mbox{\citet{2002A&A...390.1001S}}, who assume a single radiation source with a spherical density model down to scales where $T_\mathrm{dust}$~=~300~K. The inferred luminosity dominance of source~A from our work makes the single source approximation more valid than a luminosity ratio closer to unity.

When looking at Fig. \ref{fig:two_sourc_disks_torus} in Appendix \ref{app:CO}, it appears that the approximation of a single central source is not dramatically erroneous, as source~A dominates the radiation, but rather that a spherically symmetric density model breaks down in the inner regions where the temperatures exceed 90~K.
At the scales of 10$^{\prime\prime}$--15$^{\prime\prime}$ used by \mbox{\citet{2002A&A...390.1001S}}, spherically symmetrical density models are good approximations of the outer envelope of IRAS~16293 and the binary protostar is well approximated to a single source at the distances of several kAU, as well as being unresolved.
However, for radiative transfer modeling of the innermost region including the two sources, their disks as well as the conjoining dust filament, subarcsec resolution data is needed together with full 3D radiative transfer modeling to capture the complexity of the temperature structure. 
While previous radiative transfer models of IRAS~16293 do not describe the full picture, they can arguably be useful approximations to the hot corino material around source~A. If source~A is indeed dominant in terms of luminosity, then previous chemical modeling using a single star focusing on hot corino chemistry \citep{2002A&A...390.1001S, 2008A&A...488..959B, 2012A&A...539A.132C} will effectively have described the chemistry around source~A, with the hot corino chemistry around source~B acting as a "contaminant" to the signal from source~A. However, for an updated model of the hot corino chemistry involving COMs in IRAS~16293, a new chemical model is needed which includes the disks, filamentary structure and radiation fields with $L_\mathrm{A}$~$\sim$~18~L$_{\sun}$ and $L_\mathrm{B}$~$\sim$~3~L$_{\sun}$.

\section{Conclusions} 
We have constructed a 3D model of the dust and gas environment of IRAS~16293 and obtained a qualitative match to the 868 $\upmu$m dust continuum and $J$=3--2 line emission of $^{13}$CO, C$^{18}$O and C$^{17}$O from the PILS survey. This is the first model of IRAS~16293 to include multiple radiation sources as well as complex dust structures, such as disks and a dust filament. We included two different disk-like density structures; a disk structure (a true Keplerian disk) and the density solution to an infalling, rotating collapse, a disk-like structure resembling a rotating toroid. We determine that the dust density structure around source~B must be vertically extended and puffed up, as a PP-disk model around source~B with low to normal scale-heights cannot fulfill all the observational constraints. The disk structure around source~A could not be constrained. 
The results from this 3D dust and gas model suggest that the source~A luminosity is far higher than that of source~B, based on the emission morphology of the dust and CO isotopologue line emission between the sources. This model is an important increase in complexity compared to previous works and should be followed up with higher angular resolution observations of the disks and the dust filament, for further investigation and confirmation of our results. We also performed a more generic investigation of the temperature structure differences between a single star and a binary in a simple dense envelope with a density plateau within 600~AU. The main conclusions are the following:
\begin{enumerate}
  \item[--] The temperature distribution depends on the density structure and source luminosities. A single star will have higher volumes and masses of dust above $\sim$50~K compared to a binary with the same combined luminosity, while the differences in dust with 30~--~50~K are negligible in a spherical 1D density model.
      \item[--] Previous radiative transfer models of IRAS~16293 using a spherically symmetric density distribution and a single, central radiation source are valid at the $\sim$kAU scales, but not in the interbinary region, where complex 3D dust structures and two radiation sources need to be included.
      \item[--] A 3D model with disks or disk-like structures around source~A and B in IRAS~16293 and a simple dust filament between the sources, can qualitatively match the spatial distribution of the 868 $\upmu$m dust continuum and CO isotopologue line emission.
      \item[--] The density structure around source~A in our model is not constrained, as both a flat, inclined disk and an edge-on puffed-up disk structure (a rotating toroid model) can match the continuum peak emission and observed aspect ratio, while neither could match the observed C$^{17}$O isotopologue emission. 
      \item[--] The dust density structure around source~B in our model has to be vertically extended. Such a structure can be obtained with both a PP-disk model with an unusually high scale-height and with a rotating toroid model, which naturally has a puffed-up disk-like structure. 
      \item[--] The rotating toroid model abundances of the CO isotopologues need to be a factor of five lower than typical ISM values, in order to match the C$^{18}$O observations, similar to the low inferred CO abundance for other embedded protostars \citep[e.g.,][]{2016A&A...591A...3A}. This result needs to be investigated further, as the low abundances could be model dependent.
      \item[--] Source A is likely much more luminous than source~B. Our 3D density model (with coarse parameter space steps) requires the source~A luminosity to be a factor of six larger; $L_\mathrm{A} \sim 18~\mathrm{L}_{\sun}$ and $L_\mathrm{B} \sim 3~\mathrm{L}_{\sun}$.
\end{enumerate}

This work illustrates the importance of using complex, full 3D radiative transfer modeling together with subarcsecond observations of young stellar objects such as IRAS~16293 in order to constrain the physical parameters of such Class~0 disks and their environment. It also illustrates the need for even higher resolution observations of the gas and dust in the disks around source~A and B in order to do a full quantitative investigation with parameter confidence intervals, to determine the exact CO isotopologue abundances and the parameters of the dust filament enveloping both sources. We also suggest that gas dynamics, including outflows and magnetic fields, are investigated in future modeling efforts. Having a detailed and robust physical model is critical for the interpretation of ALMA spectral data of the various hot corino species.

\begin{acknowledgements}
The authors are grateful to S\o ren Frimann, Jonathan Ramsey and Attila Juhasz for useful discussions about 3D radiative transfer modeling. We would also like to thank the anonymous referee for helpful suggestions, which improved the presentation of this work.
This paper makes use of the following ALMA data: ADS/JAO.ALMA\#2013.1.00278.S. ALMA is a partnership of ESO (representing its member states), NSF (USA) and NINS (Japan), together with NRC (Canada) and NSC and ASIAA (Taiwan) and KASI (Republic of Korea), in cooperation with the Republic of Chile. The Joint ALMA Observatory is operated by ESO, AUI/NRAO and NAOJ. The group of J.K.J. acknowledges support from a Lundbeck Foundation Group Leader Fellowship as well as the European Research Council (ERC) under the European Union's Horizon 2020 research and innovation program (grant agreement No. 646908) through ERC Consolidator Grant "S4F". Research at the Centre for Star and Planet Formation is funded by the Danish National Research Foundation. M. N. Drozdovskaya acknowledges the financial support of the Center for Space and Habitability (CSH) Fellowship and the IAU Gruber Foundation Fellowship. A. Coutens postdoctoral grant is funded by the ERC Starting Grant 3DICE (grant agreement 336474).
This research made use of Astropy, a community-developed core Python package for Astronomy (Astropy Collaboration \mbox{\citeyear{2013A&A...558A..33A}}). This research has made use of NASA's Astrophysics Data System. 
\end{acknowledgements}

%
%



\bibliographystyle{aa} 

\bibliography{pap1,pap1zot} 

\appendix 

\section{SED}
\label{sec:sed}
\begin{figure*}[ht]
  \centering
      \includegraphics[width=18cm, height=18cm, keepaspectratio]{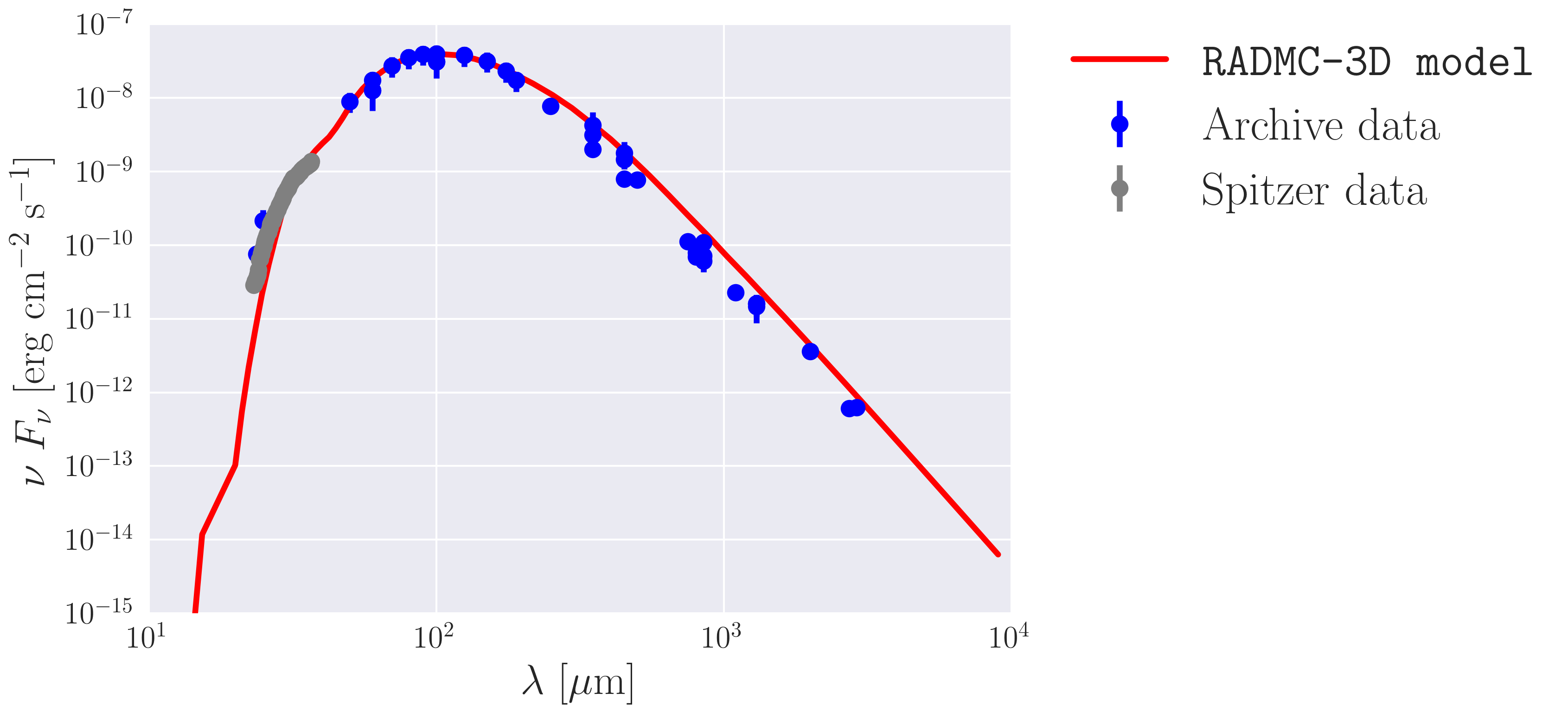}
      \caption{SED of the best rotating toroid model.}
      \label{fig:SED_1}
\end{figure*}

After the dust density components were inserted into the model, the SED was used as a sanity check that our added dust components do not alter the SED too much compared to the \mbox{\citet{2002A&A...390.1001S}} model. The envelope at $r$~>~10$^3$~AU has temperatures of 10~--~30~K, which dominates the submm and mm range, while deviations in the warmer dust mass will alter the near-infrared SED in the 10~--~100 $\upmu$m range.
In Fig. \ref{fig:SED_1} the final SED after including all structures, can be seen. While the SED appears reasonable, it should be noted that SEDs generally have very degenerate solutions, and one cannot exclude other structures or multidimensional solutions, based on the SED alone. The SEDs of the best rotating toroid model and PP-disk model were almost indistinguishable, as the SED is dominated by the cold dust in the envelope.

\section{CO isotopologue spectra}
\label{app:spectra}
\begin{figure*}
\centering
\begin{subfigure}{.33\textwidth}
\centering
      \includegraphics[height = 0.9\textheight, width=0.9\textwidth, keepaspectratio]{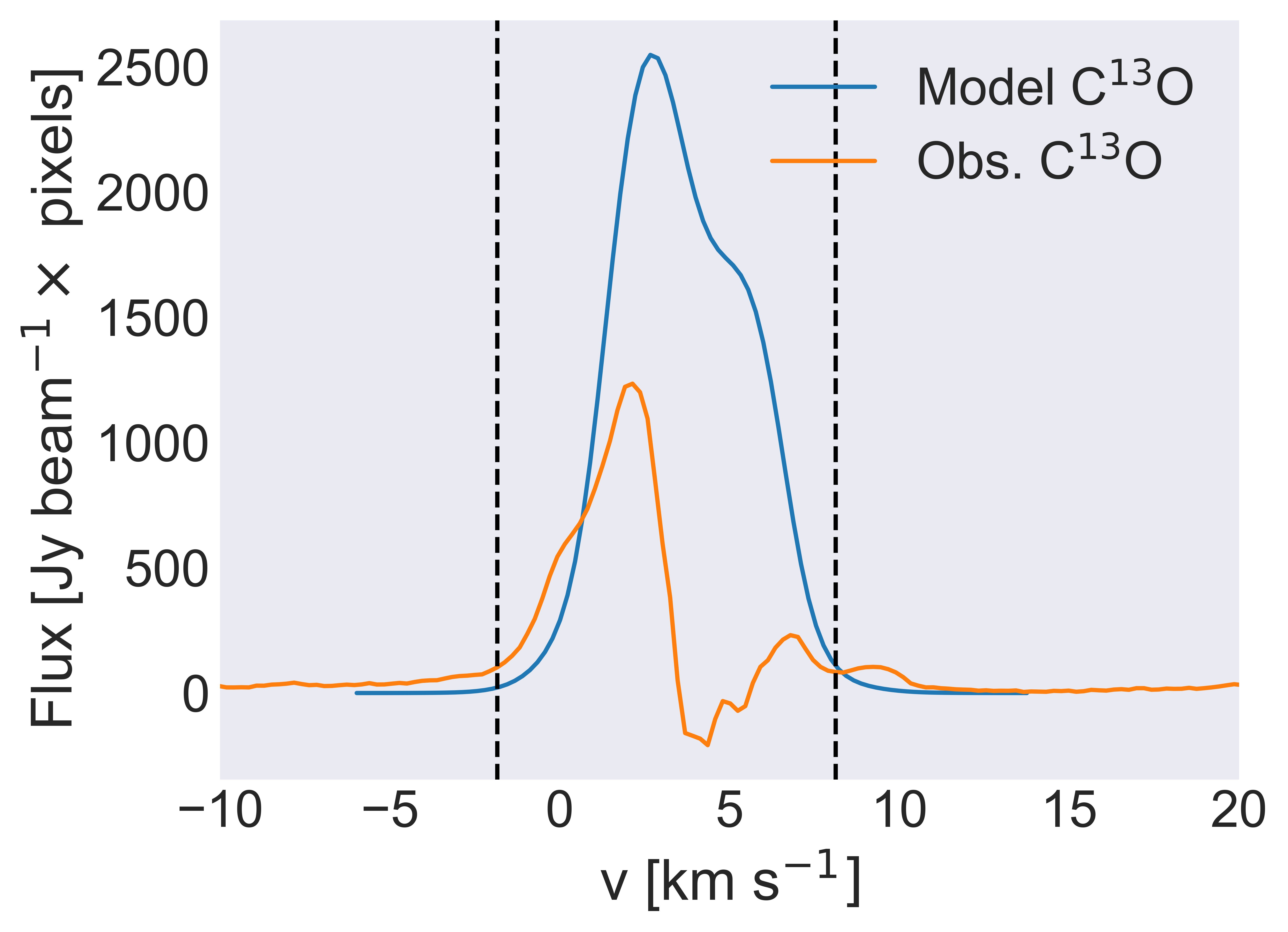}
\label{fig:test1}
\end{subfigure}
\begin{subfigure}{.33\textwidth}
\centering
      \includegraphics[height = 0.9\textheight, width=0.9\textwidth, keepaspectratio]{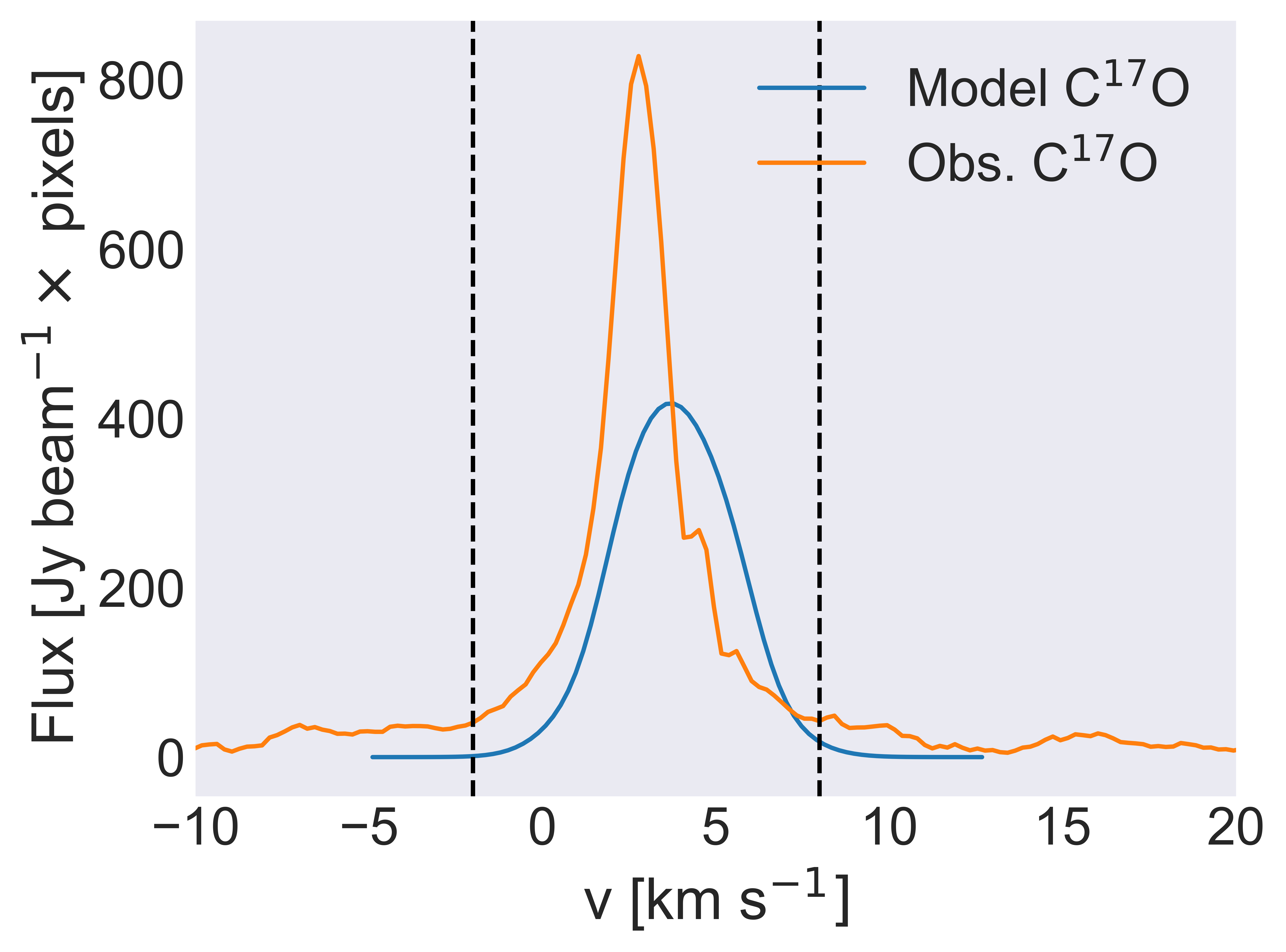}
\label{fig:test2}
\end{subfigure}
\begin{subfigure}{.33\textwidth}
\centering
      \includegraphics[height = 0.9\textheight, width=0.9\textwidth, keepaspectratio]{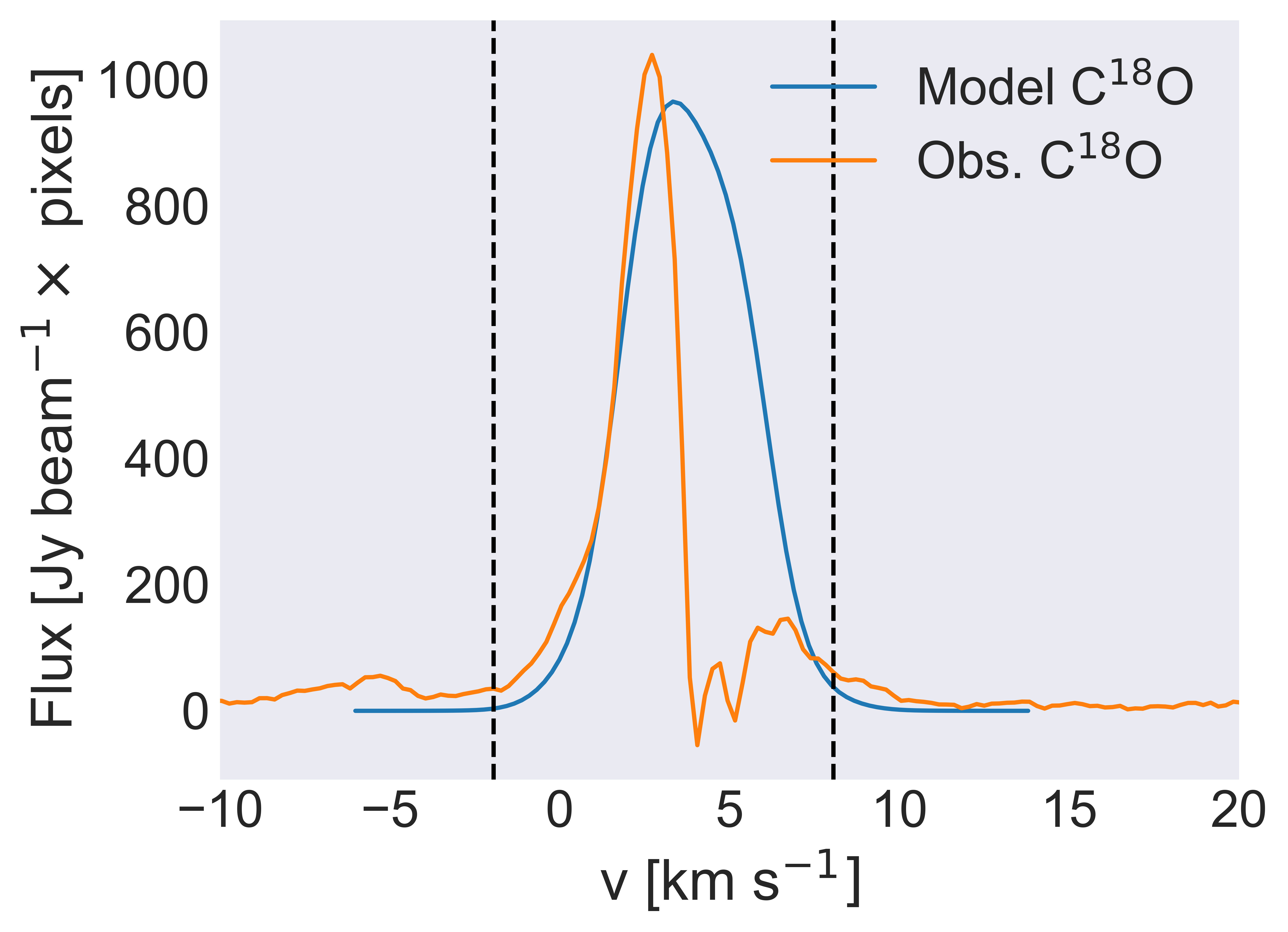}
\label{fig:test3}
\end{subfigure}
\caption{Modeled CO isotopologue spectrum and observed CO isotopologue spectrum. Vertical dashed lines show the limits of the zeroth moment maps. The model CO isotopologues are from a rotating toroid model satisfying the continuum constraints with $L_\mathrm{A}~= 14~L_\odot$ and $L_\mathrm{B}~=~7~L_\odot$.} 
\label{fig:new_spect_plt}
\end{figure*}

While we did not attempt to fit the CO isotopologues spectral lines with our models, we provide the plots in Fig. \ref{fig:new_spect_plt}, for completeness. 
A box was used around both the observations and model emission to extract the same region around the filament and two protostars, in order to avoid some of the observed extended emission south of the source~A, that we did not attempt to recreate in the models.
 
\section{Interfacing between \texttt{RADMC-3D} and \texttt{LIME}}
\label{app:radlime}
A fundamental difference between \texttt{RADMC-3D} and \texttt{LIME} is the grid used. \texttt{RADMC-3D} uses a structured grid (curvilinear is chosen for this work), while \texttt{LIME} uses Delaunay triangulation to produce an unstructured grid. Custom source code changes were made to \texttt{LIME} (v. 1.5) to allow for two local logarithmically distributed grids around the center of each of the stars in order to resolve the disks, while also changing the default density weighting value of the node positions\footnote{\url{https://lime.readthedocs.io}}, to allow the global uniform grid to resolve the dust filament, with fewer nodes, as the default setting resulted in many nodes being spread in the less interesting envelope around the disks and the filament.
The \texttt{RADMC-3D} grid density, temperature and velocity values (in the case of a rotating toroid) are written as arrays to a \texttt{C} header file, and a nearest-neighbor search of is done for each \texttt{LIME} node, which is then assigned the physical values of the nearest cell.
Due to the size of the octree refined \texttt{RADMC-3D} grid (upward of 5 million grid cells), a brute-force search is ill-advised, as this quickly becomes a speed bottleneck. Instead, a k-d tree search algorithm\footnote{The k-d tree was implemented using code modified from \url{https://rosettacode.org/wiki/K-d\_tree\#C} implemented in \texttt{C}.} \mbox{\citep[e.g.,][]{2014A&A...561A..77S}} was inserted into our custom \texttt{LIME} code. This k-d tree is used for quick lookup of the nearest neighbor, allowing us to have very large \texttt{RADMC-3D} grids without execution time issues. The exponential mode (fastest mode)\footnote{\url{https://lime.readthedocs.io/en/latest/usermanual.html\#command-line-options}} as well as parallel computing was engaged in \texttt{LIME} to reduce execution time. However, due to the large number of nodes necessary to resolve the relevant disk and filament regions, the execution time is typically around two hours per calculation, that is, six hours to produce all spectral cubes for a single model, which compounded the issue of proper parameter sampling.

\section{PP-disk model $p$ and $\Psi$ parameters}
\label{app:p_psi_par}
After running the PP-disk models over a broad parameter space of $p$, $\Psi$, $\Sigma_0$ and $H_0$ for disk~A and B (Table \ref{tab:modelsetup}), it became clear that low values of $p$ and high values of $\Psi$ result in higher peak flux densities, both in combination and by themselves, compared to the opposite part of the parameter space. A low $p$ value ensures larger amounts of dust at larger midplane radii, while a higher flaring coefficient $\Psi$ raises the dust vertically. This combination results in larger amounts of dust with relatively high temperatures.
Thus, $p$ and $\Psi$ were fixed to 0.5 and 0.25, respectively, for both disks, but neither has been constrained as optimal parameters for the two disks. However, with these fixed values, we can reproduce the maximal peak flux densities of the two disks. Another reason for these fixed values is the added computational time in \texttt{RADMC-3D} for higher $p$ values, due to higher densities in the inner region, causing significant slowdown of the photon propagation calculation, as the photon can become trapped in high-density regions, taking hundreds or thousands of absorption and reemission events to escape the cell. 
Since the goal is to see if a PP-disk model type can satisfy the constraints, but not necessarily constrain the dust distribution, this approach is acceptable. 
\section{Dust opacity in disk B}
\label{app:B_opac}
A dust opacity model could possibly accommodate the peak flux density of source~B, with lower scale-heights, making the PP-disk model more plausible for source~B. But introducing a full investigation into the dust opacity in the disks of IRAS~16293~A and B, in terms of grain size, compositions and morphology, was beyond the scope of this work and not feasible considering our computational limitations. We note that our bare grain opacity at 850~$\upmu$m is consistent with those used by \citet{2013ApJ...771...48T}, who uses 3.5~cm$^2$~g$^{-1}$ at 850~$\upmu$m, in order to fit the L1527 SED slope from 0.85~mm to 7~mm. In comparison, the bare-grain opacities that we used, which dominate the dust emission in the beam toward A and B, have a value of 3.53~cm$^2$~g$^{-1}$ at 850~$\upmu$m. As such, while unconstrained, the dust opacities used in this work are consistent with the independent modeling of another Class~0 disk.

\section{Auxiliary CO isotopologue maps and dust temperature contours}
\label{app:CO}
\begin{figure*}[ht]
    \centering
    \includegraphics[width=18cm, height=18cm, keepaspectratio]{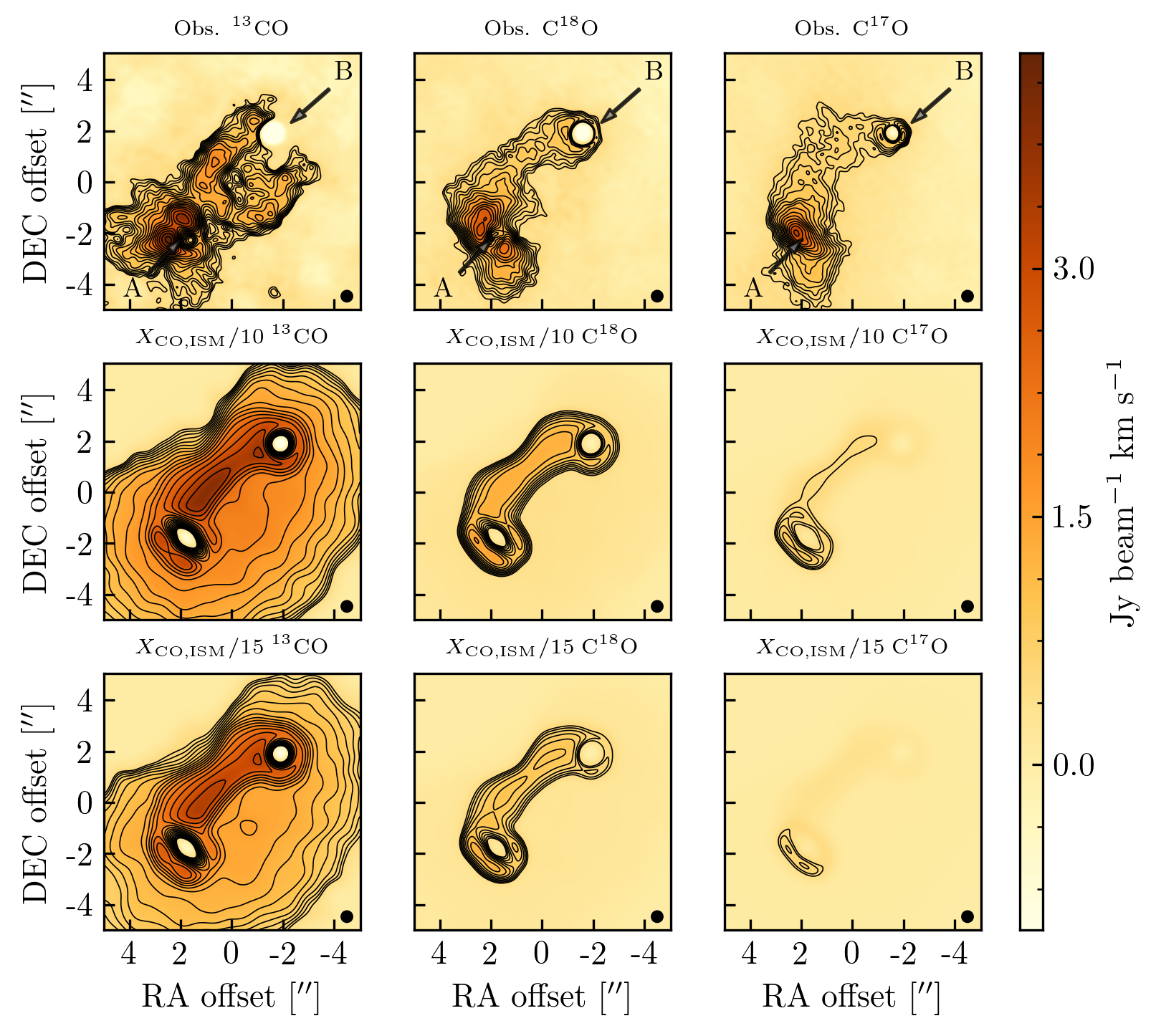}      
    \caption{Zeroth moment maps of CO isotopologue observations and synthetic gas line emission from PP-disk models. Contour levels are divided logarithmically from 0.5 to 7.3 Jy~beam$^{-1}$~km~s$^{-1}$. The RA and DEC offsets are relative to the phase center of the observations. Standard ISM abundances were divided by ten and 15, in the middle and lower panel rows, respectively.} 
    \label{fig:class_ab_divs}
\end{figure*}

Due to the absorption toward source~A in the modeled CO isotopologue maps, when using both the ISM standard abundances (Table \ref{tab:lime_mods}) and when dividing by a factor of five (Fig. \ref{fig:co_obs_mod}), the standard ISM abundances were divided by 10 and 15 in two new model CO isotopologue maps (Fig. \ref{fig:class_ab_divs}). We notice the presence of absorption in source~A and B in Fig. \ref{fig:class_ab_divs}, regardless of the CO isotopologue number density in the gas-phase, suggesting that optically thick dust is responsible for the absorption in the disk components.

\begin{figure*}[ht]
  \centering
      \includegraphics[height = 0.9\textheight, width=0.9\textwidth, keepaspectratio]{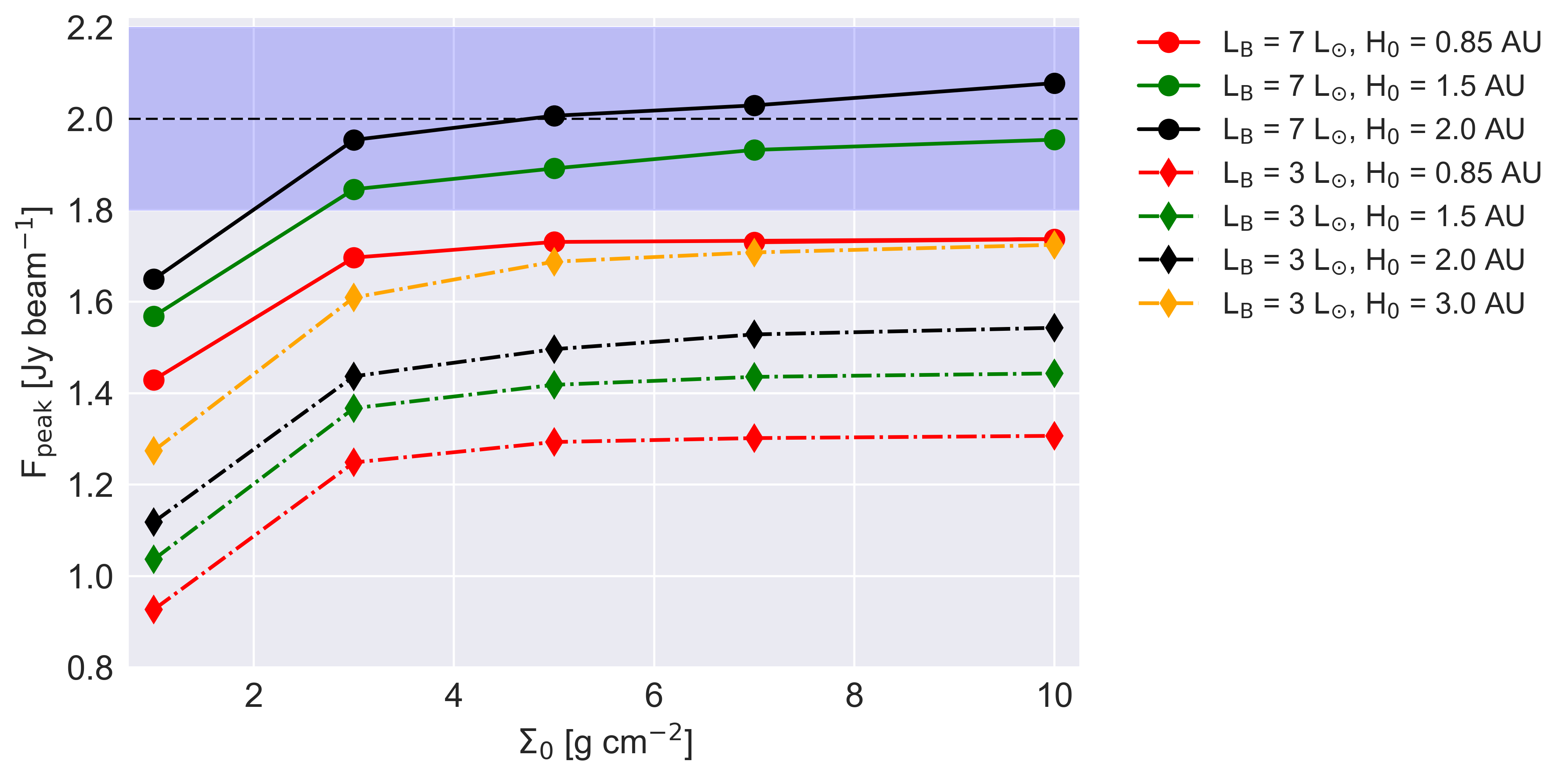}
      \caption{Peak flux density of disk~B at 868~$\upmu$m with a PP-disk model, with $L_\mathrm{B}$~=~3 L$_{\sun}$ and $L_\mathrm{B}$~=~7~L$_{\sun}$. The black dashed, horizontal line marks the peak continuum position of source~B, while the blue area represents the uncertainty. We note that the disk mass reaches 0.35~M$_{\sun}$ and 0.5~M$_{\sun}$ at 7~g~cm$^{-2}$ and 10~g~cm$^{-2}$, respectively. The highest Class~0 disk mass found in a mm survey by \mbox{\citet{2009A&A...507..861J}} is $\sim$~0.46~M$_{\sun}$, with typical Class~0 disk masses of 0.05~M$_{\sun}$.}
      \label{fig:disk_B_scale}      
\end{figure*}

\begin{figure*}
\centering
\begin{subfigure}{0.49\textwidth}
\centering
\includegraphics[width = \textwidth]{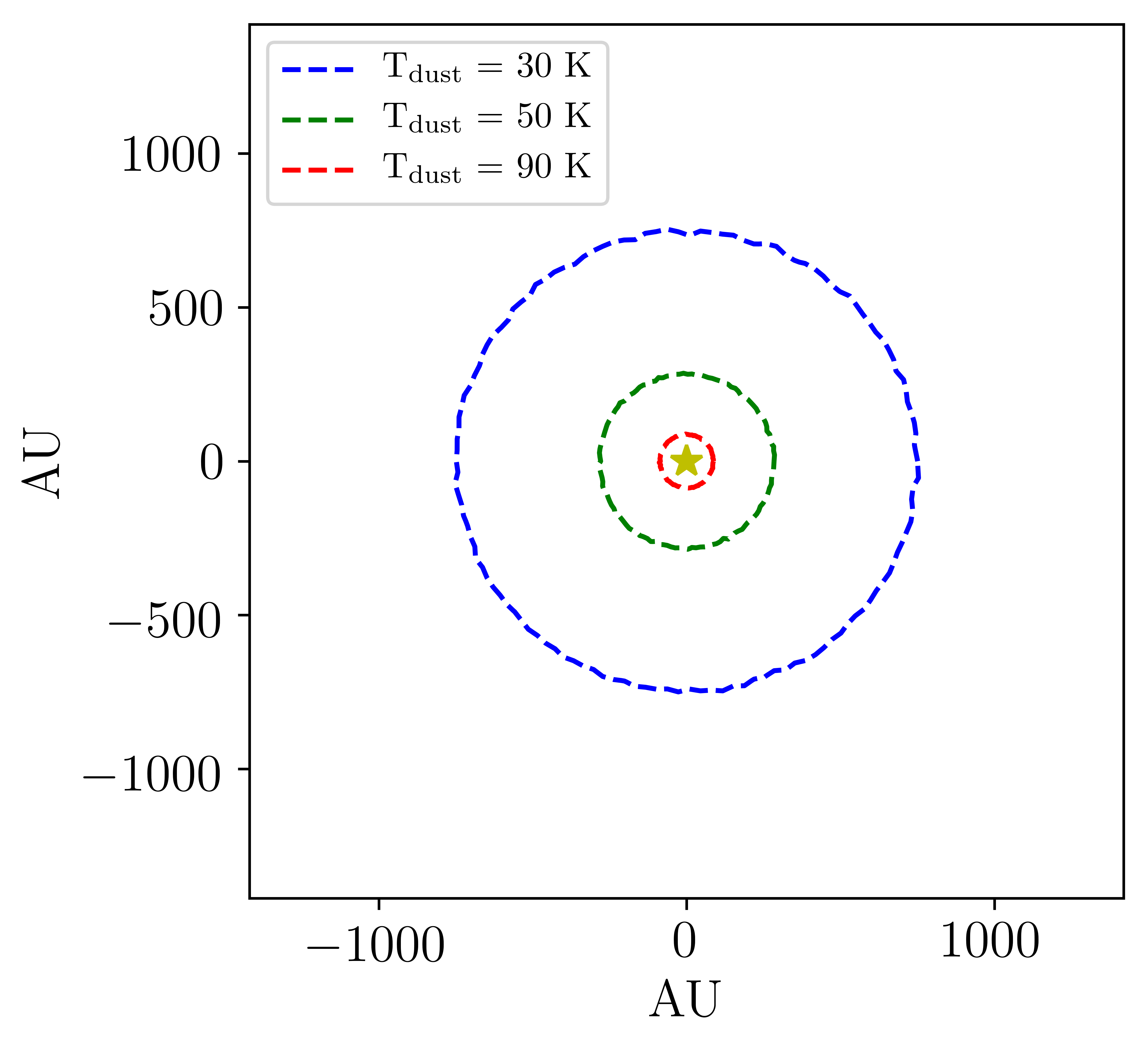}
\end{subfigure}
\begin{subfigure}{0.49\textwidth}
\centering
\includegraphics[width = \textwidth]{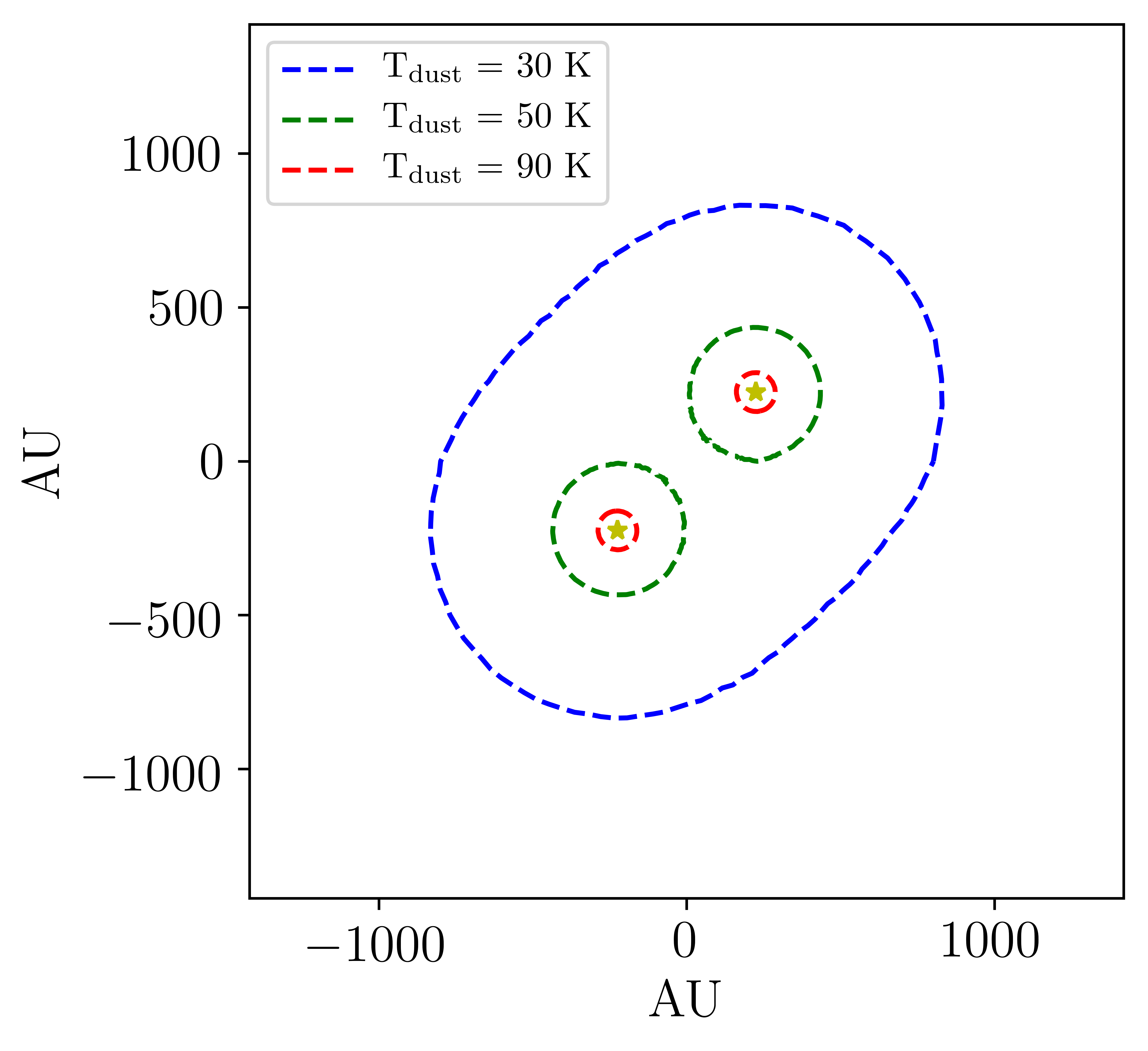}
\end{subfigure}
\caption{Temperature contours of a single star in the left panel with 21~$L_{\sun}$ in an envelope with a radial power-law density profile, $\rho_{0,\:\mathrm{env}}$ = $3.1\times10^{-14}$~g~cm$^{-3}$, $p_\mathrm{env} = 1.7$, a central density plateau within 600~AU, and two protostars in the right panel, each with 10.5~$L_{\sun}$. See Table \ref{tab:modelsetup} for more details.}
\label{fig:temp_invest}
\end{figure*}

\begin{figure*}[ht]
  \centering
      \includegraphics[height = 0.8\textheight, width=0.8\textwidth, keepaspectratio]{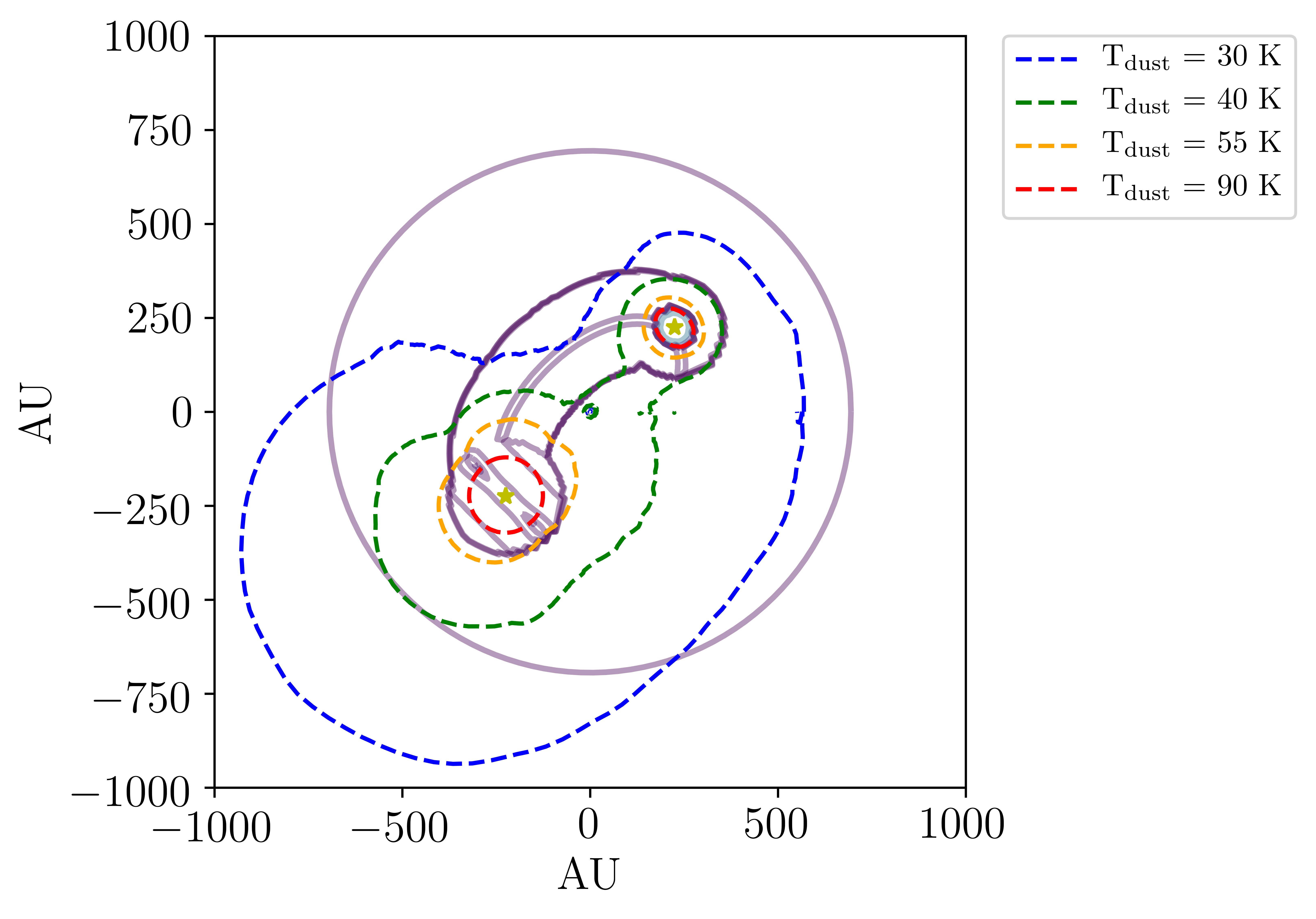}
      \caption{Temperature contours for a rotating toroid around each source and a dust filament between the sources in a slice through the model at $z$ (height) = 0. Dust density contours are shown in a gray to green colormap. The radial power-law density distribution is visible along with the dust filament and rotating toroids. Here $L_\mathrm{A}$ = 18~$L_{\sun}$ and $L_\mathrm{B}$ = 3~$L_{\sun}$. The source~A rotating toroid is edge-on while the source~B rotating toroid is face-on. The 90~K contour around source~B coincides with the transition from disk to dust filament. Due to the high number of grid cells, some temperature noise is visible in the plot center (due to very small cells with bad photon statistics in \texttt{RADMC-3D}). Interpolation has been performed along the temperature contours to make them appear more smooth for visual purposes, as the contours were otherwise slightly noisy in certain regions.}
      \label{fig:two_sourc_disks_torus}      
\end{figure*}

\end{document}